# Roadmap of spin-orbit torques

Qiming Shao, Member, IEEE, email: eeqshao@ust.hk
*Department of Electronic and Computer Engineering, Hong Kong University of Science and Technology*

Peng Li, email: peng.li@auburn.edu
*Department of Electrical and Computer Engineering, Auburn University*

Luqiao Liu, Member, IEEE, email: luqiao@mit.edu
*Electrical Engineering and Computer Science, Massachusetts Institute of Technology*

Hyunsoo Yang, email: eleyang@nus.edu.sg
*Department of Electrical and Computer Engineering, National University of Singapore*

Shunsuke Fukami, Member, IEEE, email: s-fukami@riec.tohoku.ac.jp
*Research Institute of Electrical Communication, Tohoku University*

Armin Razavi, Hao Wu, Member, IEEE, Kang Wang, Fellow, IEEE, email: arvinrazavi@ucla.edu, wuhaoiphy@gmail.com, klwang@ucla.edu
*Department of Electrical and Computer Engineering, University of California, Los Angeles*

Frank Freimuth, Yuriy Mokrousov, email: f.freimuth@fz-juelich.de, y.mokrousov@fz-juelich.de
*Forschungszentrum Juelich GmbH, University of Mainz*

Mark D. Stiles, Senior Member, IEEE, email: mark.stiles@nist.gov
*Alternative Computing Group, National Institute of Standards and Technology*

Satoru Emori, email: semori@vt.edu
*Department of Physics, Virginia Tech*

Axel Hoffmann, Fellow, IEEE, email: axelh@illinois.edu
*Department of Materials Science and Engineering, University of Illinois Urbana-Champaign*

Johan Åkerman, email: johan.akerman@physics.gu.se
*Physics Department, University of Gotherburg*

Kaushik Roy, Fellow, IEEE, email: kaushik@purdue.edu
*Department of Electrical and Computer Engineering, Purdue University*

Jian-Ping Wang, Fellow, IEEE, email: jpwang@umn.edu
*Electrical and Computer Engineering Department, University of Minnesota*

See-Hun Yang, email: seeyang@us.ibm.com
*IBM Research – Almaden*

Kevin Garello, Member, IEEE, email: kevin.garello@cea.fr
*IMEC, Leuven, Belgium; CEA-Spintec, Grenoble, France*

Wei Zhang, email: weizhang@oakland.edu
*Physics Department, Oakland University*

## Outline

## Abstract

Spin-orbit torque (SOT) is an emerging technology that enables the efficient manipulation of spintronic devices. The initial processes of interest in SOTs involved electric fields, spin-orbit coupling, conduction electron spins and magnetization. More recently interest has grown to include a variety of other processes that include phonons, magnons, or heat. Over the past decade, many materials have been explored to achieve a larger SOT efficiency. Recently, holistic design to maximize the performance of SOT devices has extended material research from a nonmagnetic layer to a magnetic layer. The rapid development of SOT has spurred a variety of SOT-based applications. In this Roadmap paper, we first review the theories of SOTs by introducing the various mechanisms thought to generate or control SOTs, such as the spin Hall effect, the Rashba-Edelstein effect, the orbital Hall effect, thermal gradients, magnons, and strain effects. Then, we discuss the materials that enable these effects, including metals, metallic alloys, topological insulators, two-dimensional materials, and complex oxides. We also discuss the important roles in SOT devices of different types of magnetic layers, such as magnetic insulators, antiferromagnets, and ferrimagnets. Afterward, we discuss device applications utilizing SOTs. We discuss and compare three-terminal and two-terminal SOT-magnetoresistive random-access memories (MRAMs); we mention various schemes to eliminate the need for an external field. We provide technological application considerations for SOT-MRAM and give perspectives on SOT-based neuromorphic devices and circuits. In addition to SOT-MRAM, we present SOT-based spintronic terahertz generators, nano-oscillators, and domain wall and skyrmion racetrack memories. This paper aims to achieve a comprehensive review of SOT theory, materials, and applications, guiding future SOT development in both the academic and industrial sectors.

# 1. Introduction

In modern computer systems, nonvolatile memories play an essential role since they can store information for a long time without the need for external power. Magnetic memories, such as hard disk drives and magnetic tapes, dominate the secondary nonvolatile memory market due to high capacity and low cost. However, they are very slow (≈ 10 ms) compared with main memories such as dynamic random-access memory (DRAM, ≈ 100 ns) and thus there is a huge speed gap. The constant data transfer between fast computing units and slow memory units creates energy loss and time delay. To mitigate this issue, high-speed nonvolatile memories such as flash memory (≈ 0.1 ms) have been employed for personal and enterprise electronic devices. In addition, emerging nonvolatile memories, including but not limited to resistive RAM (RRAM), phase change memory, magnetic race-track memory, and magneto-resistive RAM (MRAM), are being intensively investigated to further reduce the gap [1], [2]. The concepts behind magnetic race-track memory are close to those behind magnetic bubble memory, which was developed in the 70s and 80s, except that magnetic race-track memory was originally based on electrical manipulation of the magnetization using spin-transfer torque (STT) [2]. Similarly, while the first generation of MRAM used current-induced Oersted fields to write information, which does not scale well, STT-MRAMs use currents directly [3]–[5]. STT-MRAM is being actively investigated for both embedded and standalone applications by several prominent industry leaders. Thanks to its ultrafast speed and reliability, STT-MRAM could potentially achieve cache-level speed (≈ 1 ns), resulting in increased interest in STT-MRAM [6]. In 2011-2012, a new type of current-induced spin torque, spin-orbit torque (SOT), was employed to enable a new type of MRAM, known as SOT-MRAM [7], [8]. The SOT writing scheme can provide better energy efficiency and provides more application versatility. To fulfill its potential, device characteristics including the tunnel magnetoresistance (TMR) ratio, back-end-of–line (BEOL) process compatibility, write efficiency, storage density, and circuit level design and integration need to be further improved (Fig. 1), as discussed further in section 4h. We discuss the importance of these criteria here. TMR ratio determines how effectively one can read the MTJ state and minimize the reading margin. Larger TMR ratio (targeting toward 250 %) provides a larger reading margin, enabling a faster reading speed. It has also a significant impact on the periphery area, which can occupy a significant portion of chip area, hence affecting effective memory density. Further, to ensure that SOT-MRAM is appropriate for memory applications, large enough MTJ's thermal stability $\Delta$ at small diameters (30 nm) is critical for achieving sufficient retention time at operation temperature (typically $\Delta > 45\ k_bT$). Meanwhile, the MTJ stack has to sustain significant processing thermal budgets (~400 °C, >30 min) without degrading its properties to be compatible with CMOS back-end-of-line technology integration and processing.

Importantly, while SOT-MRAM does not consume static power due to its nonvolatile nature, reducing dynamic power is still critical to minimize cell footprint (selector transistor size, write latency). This calls for reduction of writing efficiency, and it is projected that one would need to improve the SOT efficiency to a value larger than one so that the switching current can reach write latency 1 ns with <100 μA in advanced technology nodes. Providing sufficient advantages compared to SRAM memories (primary technology replacement target) requires that the area cost of SOT-MRAM is less than that of Static RAM (SRAM) and that eventually the density of SOT-MRAM should approach that of STT-MRAM. As density strongly impacts yield, the fabrication process needs to be optimized, requiring strong effort on etch methods to avoid vertical shorts and MTJ damage at lower dimensions (30 nm).

In the following text, we will briefly introduce SOT and its mechanisms, and explain why it can be potentially more energy-efficient and versatile in its applications.

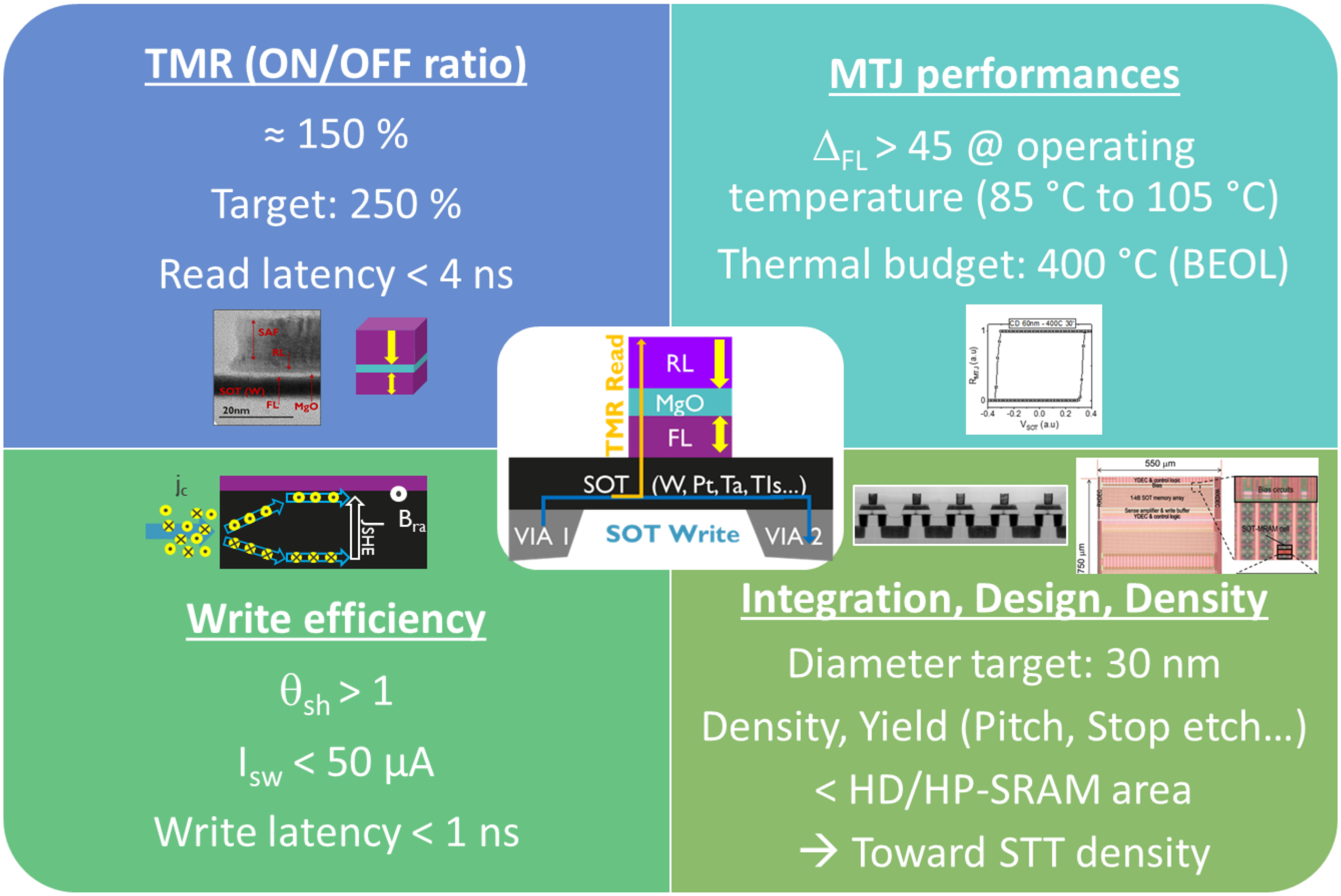

**Figure 1.** Areas of future development and targets for SOT-MRAM. MTJ: magnetic tunnel junction; $\Delta_{FL}$, thermal stability factor of the magnetic free layer; $\theta_{SOT}$: SOT efficiency; $I_{sw}$: critical switching current; HD/HP-SRAM: high-density/high-performance static random-access memory; STT-MRAM: spin-transfer torque-magnetoresistive RAM; RL: reference layer; FL: free layer; TIs: topological insulators. Insets: TEM-array: ref. [9] ©[2018] IEEE, SOT-MRAM-array: ref. [10] ©[2020] IEEE , SOT-MRAM circuit: ref. [11] ©[2020] IEEE.

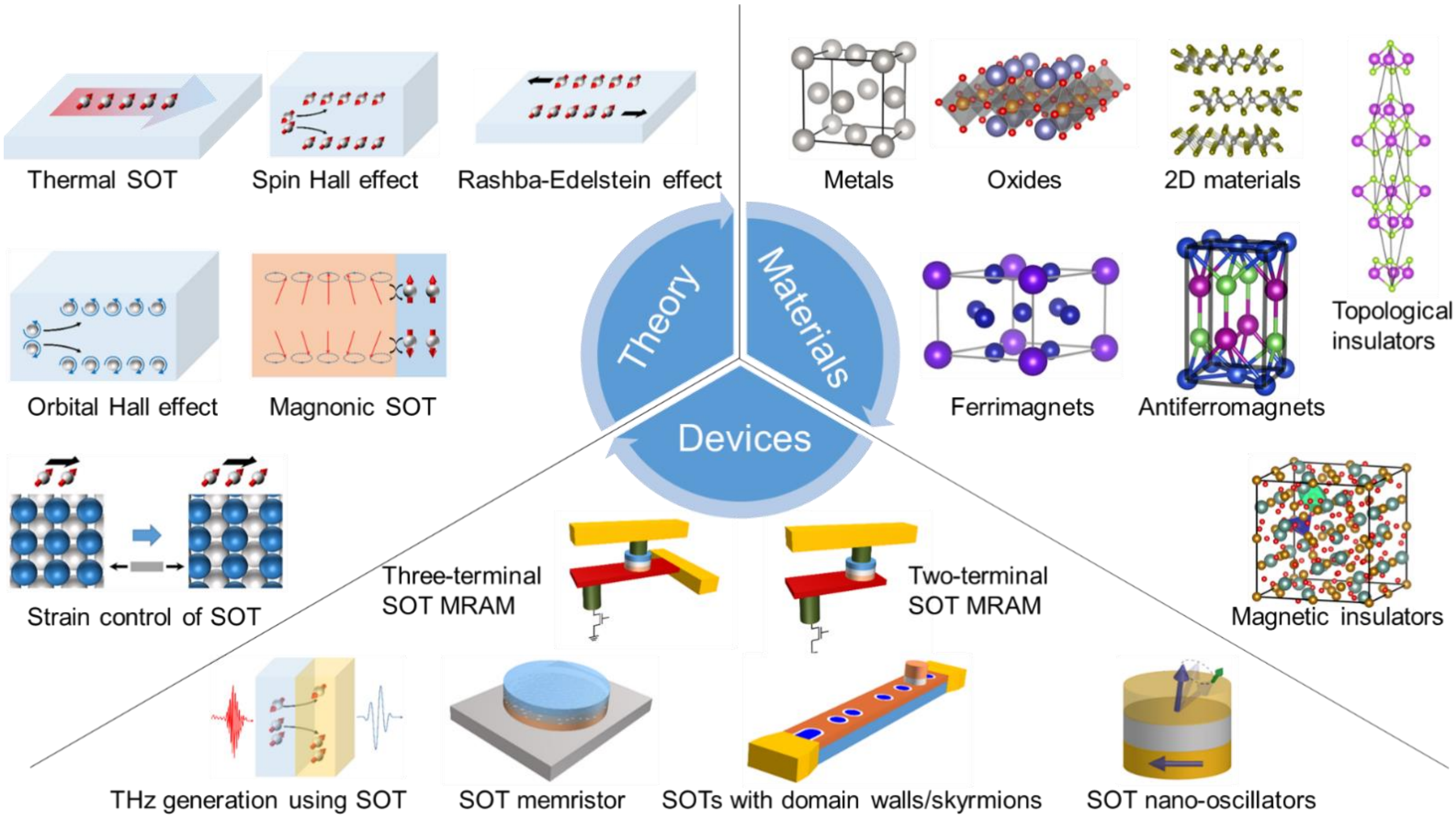

**Figure 2**. Overview of the roadmap content. The spin Hall effect and Rashba-Edelstein effects are introduced in section 2a; the orbital Hall effect is introduced in section 2c; thermal SOTs are introduced in section 2d; strain control of SOT is introduced in section 2e; magnonic SOTs are introduced in section 2f; metals are discussed in section 3a; topological insulators are discussed in section 3b; 2D materials are discussed in section 3c; oxides and magnetic insulators are discussed in section 3d; antiferromagnets are discussed in section 3e; ferrimagnets are discussed in section 3f; three-terminal SOT-MRAM and two-terminal SOT-MRAM are elaborated in section 4a and 4b, respectively; SOT memristors are elaborated in section 4c;THz generation using SOT is elaborated in section 4e; SOT nano-oscillators are elaborated in section 4f; and SOTs with domain walls/skyrmions are elaborated in section 4g. In addition to the topics mentioned in the figure, we discuss first-principle calculations in section 2b, low-damping ferromagnets in section 3g, schemes to achieve field-free magnetization switching in section 4d, and industrialization considerations in section 4h.

SOT is generated from spin-orbit coupling in a single material or material heterostructure. This is distinctively different from STT, where the spin angular momentum is transferred from one magnet to another magnet. Conventionally, an electric field creates nonequilibrium orbital occupation that leads to SOT through spin-orbit coupling. This process includes the spin Hall effect [12], [13], the Rashba-Edelstein effect [14], [15], and the orbital Hall effect [16], [17]. More recently, other types of spin torques, such as thermally generated or phonon-driven spin torque [18], [19] and magnon-driven spin torque [20], [21], have been studied as SOTs. Investigations into mechanisms of SOTs have allowed researchers to control SOT strength through gate voltage [22], strain [23], [24], and other external parameters.

SOT is attractive because of its potential energy efficiency. The efficiency for generating spin current and transferring angular momentum can be represented by a dimensionless number that characterizes the angular momentum transferred per electron. The STT efficiency has a theoretical limit of one since it originates from the spin polarization at the Fermi level of a ferromagnetic metal, which has a maximum of 100 %. In contrast, SOT efficiencies larger than one have been experimentally demonstrated in some topological materials and Rashba heterostructures. Since the required switching current is inversely proportional to the STT or SOT efficiency, the energy efficiency of SOT could be better. Researchers are investigating many materials for higher energy efficiency, including heavy metals, topological insulators, 2D materials, and oxides. In addition, ferromagnetic, ferrimagnetic, and antiferromagnetic materials are also studied for generating SOTs, in concert with the studies of magnetism and magnetic dynamics in the magnetic materials.

SOT-based devices (e.g., SOT memory) usually have separate write and read paths. This increases the reliability of the devices since one does not need to apply a large write current through the ultrathin (≈ 1 nm) tunnel dielectric layer in a magnetic tunnel junction. Furthermore, in some applications, SOT devices do not require a second magnetic layer as in the STT case, making pure spin-based logic and computing units more tangible and providing more flexibility in terms of device structure design. Potential applications include, but are not limited to, MRAM, neuromorphic circuits, race-track memory, nano-oscillators, and terahertz generation. Early reviews of SOT-based devices include refs. [25]–[28].

In this paper, we review and provide a perspective of SOTs in both the theoretical and experimental aspects of their origins, materials and devices. We begin with the theory of SOT. We introduce various mechanisms to generate SOTs, including spin Hall effect, Rashba-Edelstein effect, orbital Hall effect, strain effect, thermal effect, and magnonic effect. Insights from first-principle calculations are briefly mentioned. Then, we describe materials for generating SOTs including metals and metal alloys, topological insulators, 2D materials, and oxides. Then, we discuss SOTs from magnetic materials, including ferromagnets, antiferromagnets, and ferrimagnets. Finally, we discuss SOT devices. We introduce three-terminal and two-terminal SOT-MRAM, SOT-based neuromorphic devices, nano-oscillators, race-track memory with domain walls and skyrmions, and terahertz generators based on SOTs. We finish with a discussion of a variety of mechanisms for predictably switching SOT devices without the assistance of external fields, a requirement for scaling SOT-based devices.

## 2. Theory of spin-orbit torques

### a. Spin Hall effect and Rashba-Edelstein effect as origins of SOTs

Two simple models provide the paradigms for understanding SOTs. The first one is that SOTs can arise from the interaction of spin-orbit coupling and magnetic exchange at an interface between a magnetic layer and a nonmagnetic layer, and has been proposed by Manchon and Zhang [29] based on the Rashba-Edelstein effect [14]. In a two-dimensional electron gas (2DEG) with magnetic exchange and spin-orbit coupling, a current combines with the spin-orbit coupling to create a spin accumulation that exerts a torque on a magnet (Fig. 3a). Note that the interfacial SOT magnitude would be dependent on the extrinsic factors, such as the disorder strength [30], [31]. The second model is SOTs arise in a bi-layer structure of a non-magnetic layer and a ferromagnetic layer, from spin currents generated in the non-magnetic layer that couple to the ferromagnetic layer at the interface [32], [33] based on the spin Hall effect [13]. Note that the symmetry of these intrinsic SOTs can be directly related to the crystal symmetry, which for example, can be Rashba-like [14], [34] or Dresselhaus-like [34], [35]. Here, the spin Hall effect generates the spin current that flows to the interface, creating a spin torque similar to that created by a perpendicular spin-polarized current (Fig. 3b).

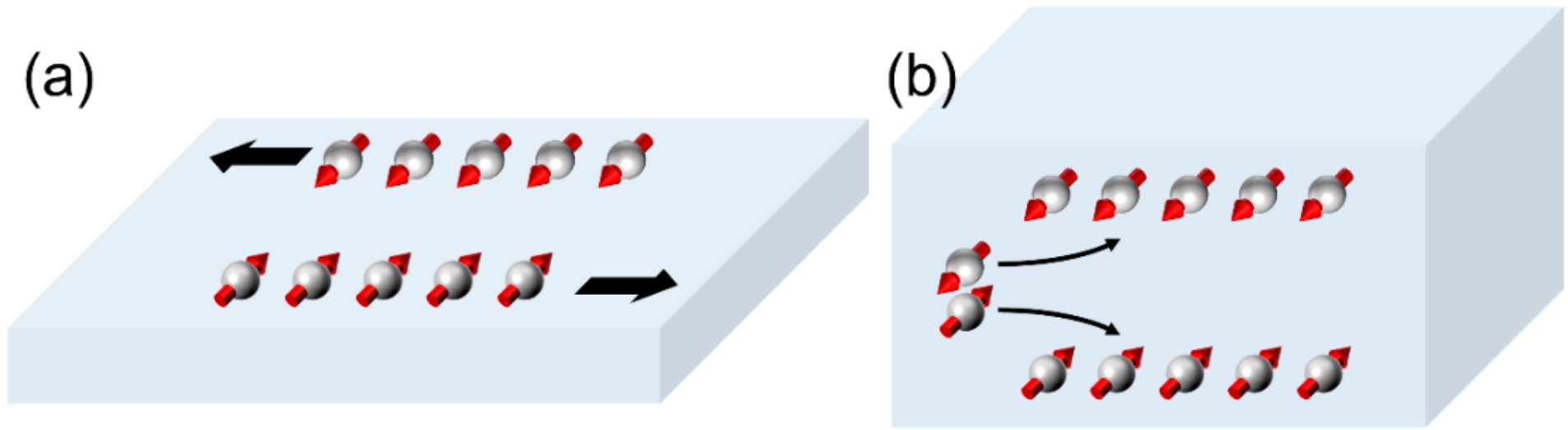

**Figure 3**. Schematics of (a) Rashba-Edelstein effect and (b) spin Hall effect

Models [36] that include both the spin Hall effect and interfacial spin-orbit coupling show that both paradigms produce both fieldlike $\sim\boldsymbol{m}\times(\boldsymbol{n}\times\boldsymbol{E})$ and dampinglike $\sim\boldsymbol{m}\times[\boldsymbol{m}\times(\boldsymbol{n}\times\boldsymbol{E})]$ torques, where $\boldsymbol{m}$ is the magnetization direction, $\boldsymbol{n}$ is the interface normal direction, and $\boldsymbol{E}$ is the direction of the electric field. It is impossible to use the form of the torque to determine the mechanism since both mechanisms give rise to the same form of the torque. If the materials in the bilayer are assumed to have isotropic symmetry, as might be appropriate for polycrystalline materials, these two forms are the simplest allowed by symmetry [37].

Theoretical research on SOTs has focused on developing a deeper understanding beyond these two paradigms to see where additional functionality can be achieved and then optimized. One approach is to study the angular dependence of the torque in more detail. For isotropic materials, SOTs with more complicated angular dependencies than the two described above are allowed [37], [38]. However, permitted torques for isotropic materials all go to zero when the magnetization points in the direction $\boldsymbol{n}\times\boldsymbol{E}$. This makes it impossible to deterministically reverse magnetizations with perpendicular anisotropy, that is, for the easy axis along the interface normal $\boldsymbol{n}$. Deterministic switching requires that something break the mirror symmetry with respect to the plane perpendicular to $\boldsymbol{n}\times\boldsymbol{E}$. The symmetry breaking can come from a variety of sources including exchange bias, an applied magnetic field, or the crystal symmetry [38], [39]. MacNeill et al. [39] used a single crystal $WTe_2$ layer and observed such a torque when the electric field was applied in the direction such that the crystal lattice broke the mirror plane perpendicular to $\boldsymbol{n}\times\boldsymbol{E}$, but not when the lattice is oriented so that it did not. One avenue for future research is to identify crystalline materials that can produce large torques for magnetizations along the $\boldsymbol{n}\times\boldsymbol{E}$ direction.

Another way to break the necessary mirror plane is with an additional ferromagnetic layer that has a magnetization that points in a different direction than the magnetization of interest. For example in a trilayer system with two ferromagnetic layers, the anomalous Hall effect and the anisotropic magnetoresistance in the second ferromagnetic layer can generate spin currents that reach the layer of interest and that have spins oriented in such a way as to generate torques with novel angular dependencies [40]. In addition, spin currents due to the spin Hall effect and spin currents from spin-polarized ferromagnetic transport can be rotated at the interface between the second ferromagnetic layer and the spacer layer. These rotated spin currents can also create torques with novel angular dependencies [41][42]. Switching due to torques from a second ferromagnetic layer has been demonstrated experimentally [43]. Optimizing the torques from such systems forms another important area of future research. The interfaces of ferromagnetic layers not only can create spin currents that can create torques on other layers, they can also create self-torques on that same layer [44]. In symmetric layers, the torques on both interfaces cancel each other as a net torque but can lead to observable rotations of the magnetization near the interface [44]. An interesting area of research is whether making the two neighboring layers sufficiently different can lead to useful net torques on the magnetization.

## b. First principle calculations

First-principles electronic structure calculations are playing an increasing role in our understanding of SOTs, and that role can be expected to continue to increase. Their impact has been delayed because interpreting the results of such calculations is complicated. One complication is that all mechanisms are mixed together and need to be untangled. One approach for untangling them is to modify parts of the energies, for example by artificially changing the spin-orbit coupling in different regions, to see which region plays an essential role [36][45]–[47]. Another complication is the unknown role played by disorder. The bilayers in which SOTs are large tend to have very poor lattice matching between the materials, the materials are typically polycrystalline, and they tend to be quite resistive. Unfortunately, the detailed structure of the interfaces is not typically characterized. Recent calculations have attempted to capture the high resistivity by including a random site potential [38][48], [49]. This approach addresses some of the complications but not ones like lattice mismatch and polycrystallinity that would require prohibitively large unit cells. Future advances in first-principles calculations will enable the treatment of more realistic structures.

## c. Orbital Hall effect-induced SOTs

The perturbation that generates spin currents in these bilayers, the electric field, does not couple directly to the spin but rather to the orbital degrees of freedom. Even without the spin-orbit coupling needed for spin currents, a transverse orbital angular momentum current is possible (Fig. 4), raising the question of whether orbital currents may play a fundamental role in these systems. Bulk calculations of the orbital Hall current suggest that it can be large [16], [17]. However, since orbital moments couple strongly to the lattice and spins do not, when an orbital current reaches the interface with a ferromagnet it is not clear whether it transfers angular momentum through the spin-orbit coupling to the magnetization or directly to the lattice. Calculations [50] suggest that it can exert a torque on the magnetization, particularly in bilayers in which the spin-orbit coupling is weak in the non-magnet and strong in the ferromagnet. According to the calculations, the torque from orbital-current injection has the same field-like and damping-like components as the torque from the spin current injection. In general, the spin injection and orbital injection torques compete with each other. Thus, the sign and magnitude of the net torque acting on the magnetization can be different from that predicted based only on the spin injection mechanism [51]. This suggests a route to enhance the torque efficiency in spintronic devices: identifying systems in which the orbital injection and spin injection torques have the same sign. In (TmIG)/Pt/CuOx the Pt/CuOx interface is the source of a large orbital current, which is converted into a spin current in Pt and then injected into TmIG, where it exerts a torque on the magnetization [52]. This additional torque may increase the SOT efficiency in this device by at most a factor of 16. The potential role of orbital currents in switching devices remains an outstanding question for theory, experiment, and potential applications.

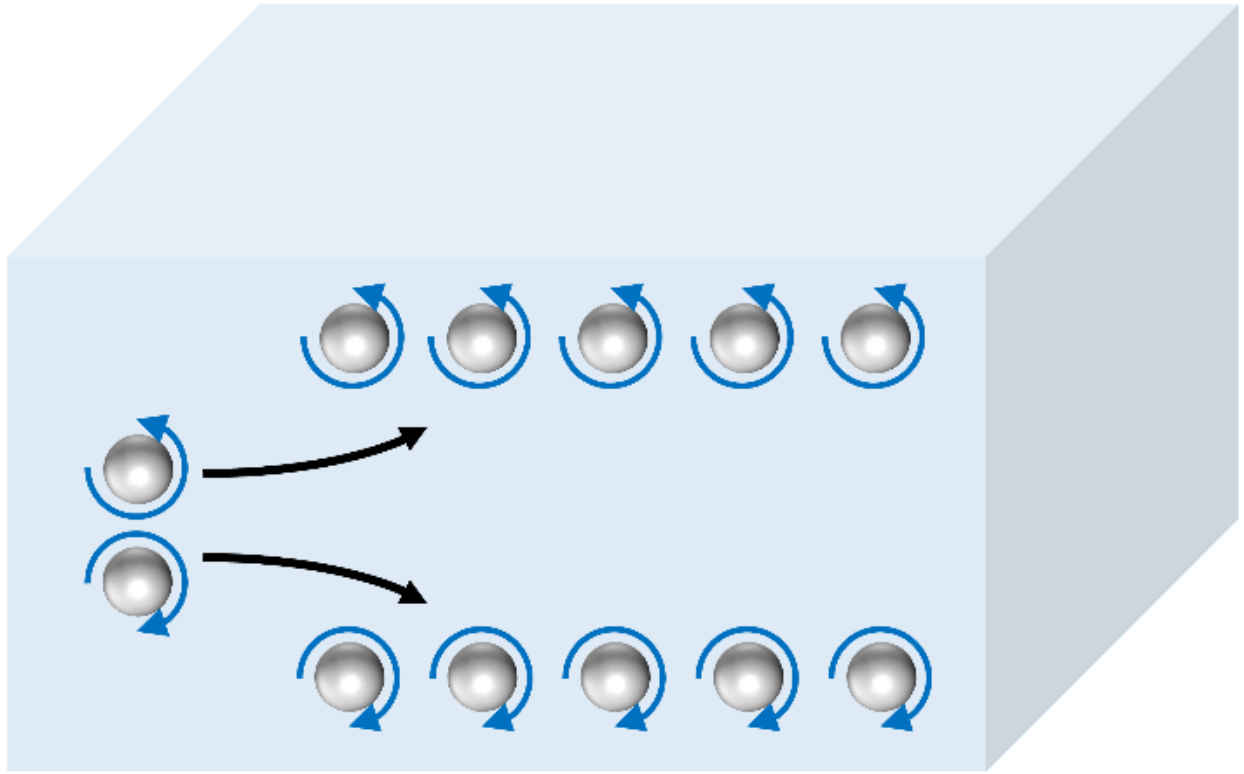

**Figure 4**. Schematic of orbital Hall effect, where circulating blue arrows indicate the orbital angular momentum direction.

## d. Thermal generation of SOTs

The interplay between thermal effects and spintronic effects leads to an emerging area of spin caloritronics [53], [54]. One aspect of SOTs that is starting to be investigated is the use of thermal gradients to drive magnetization dynamics in ferromagnets and antiferromagnets. For example, the thermal analogs of the effects of SOTs and inverse SOTs [18], [19]. This area combines SOTs with modern spin caloritronics [55], which to date has focused on utilizing thermal gradients instead of electric fields to drive spin currents, studying phenomena such as the spin Seebeck, anomalous Nernst, and the spin Nernst effects. Recent calculations suggest that the application of a temperature gradient can exert a torque on the magnetization (Fig. 5), that is, a thermal SOT [19], and, conversely, the inverse thermal SOT can be responsible for magnetization dynamics driving heat currents [18].

In analogy to the SOT driven by electrical currents, thermal SOTs can be decomposed into even and odd components concerning the magnetization direction, and they have the same symmetry properties. The intrinsic even part of the thermal SOT is analogous to the intrinsic anomalous Nernst [53] and spin Nernst effects [56], of which the latter can also be identified as the source of the even thermal SOT in magnetic bilayers. The thermal SOT can be computed directly from its electrical counterpart employing a Mott-like relation [18], [19]. It can be made as large as the electrical SOT by proper electronic structure engineering. Preliminary calculations show that fast domain walls moving at a speed of the order of 100 m/s can be used to drive significant heat currents via the inverse thermal SOT [18], [19]. Experimentally, the thermal SOT has been measured in W/CoFeB/MgO and it has been shown to assist the electrical SOT by lowering the critical current needed for magnetization switching in this material [57]. In addition, thermal SOT-induced magnetization auto-oscillation has been experimentally demonstrated [58].

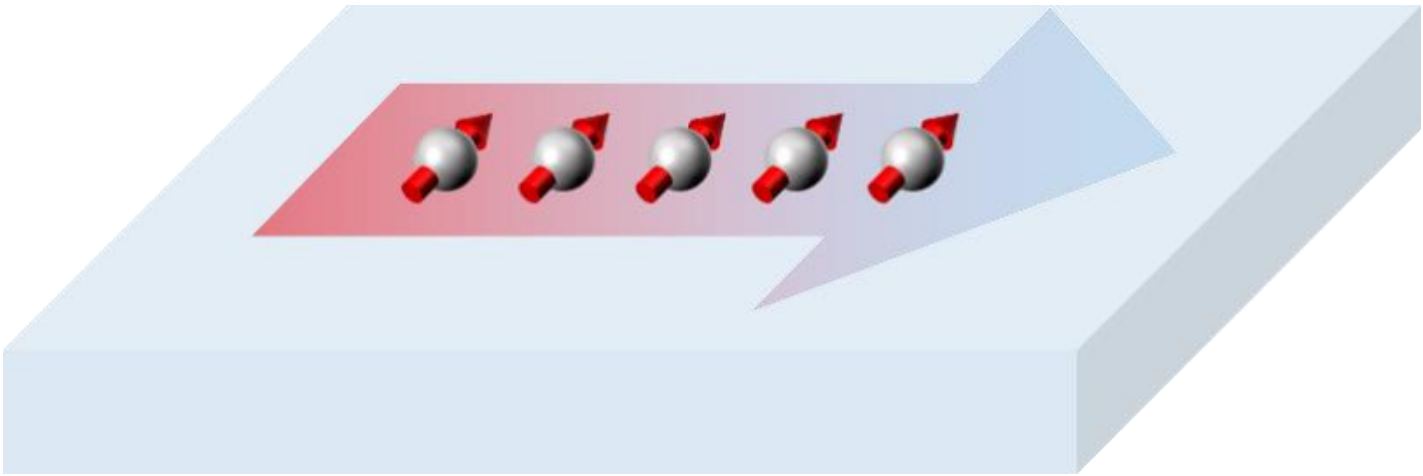

**Figure 5**. Schematic of thermal gradient-induced spin current.

## e. Strain control of SOTs

For many applications, it would be desirable to control the SOT with external means in addition to the current. While sample parameters such as composition and layer thickness of the ferromagnet/heavy metal heterostructures can be adjusted to design the magnitude and the sign of SOTs, they do not allow "dynamical" control in a given system on-demand. One energy-efficient tool for that is electric-field driven mechanical strain [59]. Strain is an efficient means to tune magnetic properties such as the magnetic anisotropy energy without ohmic losses. It can be applied locally, which is a key element for developing and realizing complex switching concepts in novel devices. The first attempts to investigate the effect of strain on switching by spin torques focused on the effect of strain on the anisotropy and the resulting impact on the switching in systems with an in-plane magnetic easy axis [60], [61].

Many proposed applications of SOTs are based on systems with perpendicular magnetic anisotropy rather than in-plane magnetized materials. Recent experiments demonstrate electrically induced strain control of SOTs in perpendicularly magnetized W/CoFeB/MgO multilayers grown on a piezoelectric substrate [24]. The experiments show that the strain, as modulated by the electric field applied across the piezoelectric substrate, leads to distinct responses for both the field-like and damping-like torques, with a factor-of-two change (Fig. 6). Ab-initio calculations of the SOT within the Kubo formalism performed for FeCo alloys on W(001) reveal that this happens due to redistribution of *d*-states upon the reduction of symmetry as the strain is introduced. Some of the *d*-states that mediate the hybridization with the heavy-metal substrate transform differently with respect to tensile or compressive strain. As the field-like and damping-like SOTs originate from different electronic states, they generally follow distinct dependencies on structural details. In this case, the field-like term is hardly affected by the strain, but the magnitude of the damping-like component changes drastically, increasing by as much as 35 % as the lattice is expanded by 1 % along the electric-field direction. These findings integrate two energy-efficient spin manipulation approaches, electric field-induced strain, and current-induced magnetization switching, establishing a new research direction for the development of novel devices.

Recently, electric field control of SOTs in antiferromagnetic heterostructures, $Mn_2Au$/ $Pb(Mg_{1/3}Nb_{2/3})_{0.7}Ti_{0.3}O_3$ (PMN-PT), has been experimentally demonstrated [62]. Also, Cai et al. achieved field-free SOT switching in a ferroelectric/ferromagnetic structure by applying a lateral electric field on PMN-PT [23].

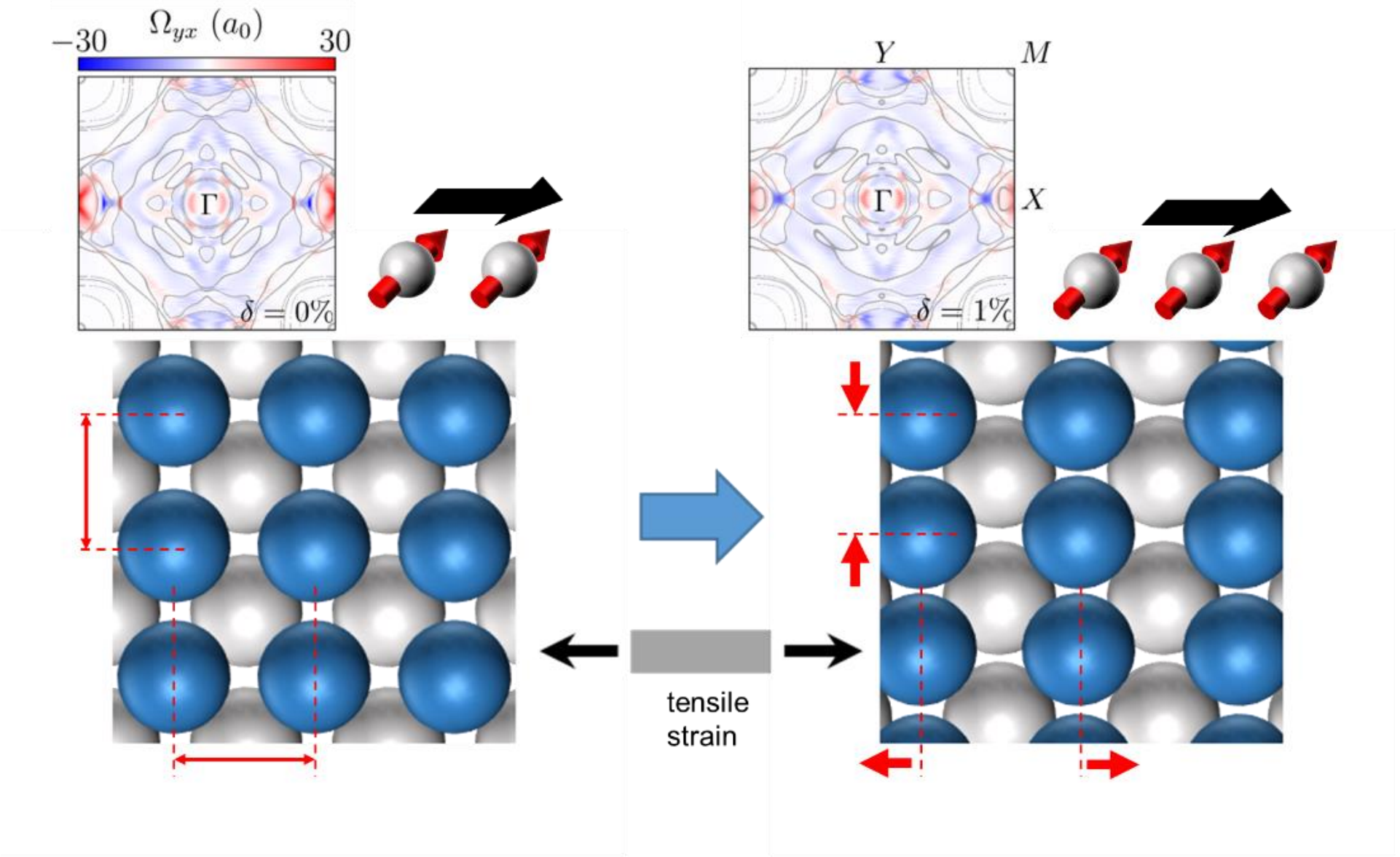

**Figure 6**. Strain control of SOT, where an increase of a tensile strain can lead to a change of spin density of states and thus an increase of SOT strength. δ is the fractional strain. The upper panel shows the microscopic contribution of all occupied bands to the damping-like spin-orbit torque ($\Omega_{yx}$) in relaxed and strained crystal structures. Gray lines indicate the Fermi surface. The upper panel is reprinted with permission from [24]. Copyright (2020) by the American Physical Society

## f. Magnonic SOTs

The interplay between thermal effects and spintronic effects leads to an emerging area of spin caloritronics [53], [54]. Out of the many activities in modern spintronics, an important milestone is the discovery of electrical spin-torque associated with electron-mediated spin currents [63]–[65]. This breakthrough has opened the era for electrically controlled magnetic device applications, for instance, magnetic random-access memories. However, the electron-mediated spin torque, involving moving charges, suffers from unavoidable Joule heating and corresponding power dissipation, as well as a short spin propagation length [66]. These fundamental obstacles can be overcome by magnon-mediated spin-torque, in which the angular momentum is carried by precessing spins rather than moving electrons [67]. Therefore, much less Joule heating occurs. Moreover, magnon currents can flow over distances of micrometers [68]–[71] even in insulators, and thus, materials are not limited to electrical conductors. The readers can refer to recent reviews [67], [72] for details about magnon spintronics. Most magnon-related magnon-related studies have been focused on long-distance magnon transport. Recently, magnon-torque-driven full magnetization switching was achieved through an antiferromagnetic insulator NiO by injecting an electric current to an adjacent topological insulator $Bi_2Se_3$ [20]. While the magnon torque ratio of 0.3 is larger than the electrical spin-torque ratio in heavy metals and is comparable to that of topological insulators reported recently [73], [74], the existence of the coherent antiferromagnetic magnon media such as 25 nm thick NiO reduces the spin-torque ratio from 0.67 to 0.3 [20]. Therefore, research works need to be carried out in order to eliminate such a drop in the spin-torque ratio due to magnon transport. For SOT-MRAM applications, increasing the magnon torque ratio > 0.8 is required to achieve sub 100 µA current operation. So far, the experiments rely on the interconversion between magnon spin current and electron spin current at the magnetic insulator/heavy metal interfaces (Fig. 7). Pure magnon-driven magnetization switching with no electrical contributions may be realized in the near future. In addition, the demonstrated magnon-torque scheme provides a solution for exploring magnetic devices based on newly discovered quantum materials, in which the issue of current shunting can be eliminated by electrically separating the current path from the magnetic layer. Furthermore, the electrical isolation relaxes the requirement for heterogeneous integration with a magnetic layer.

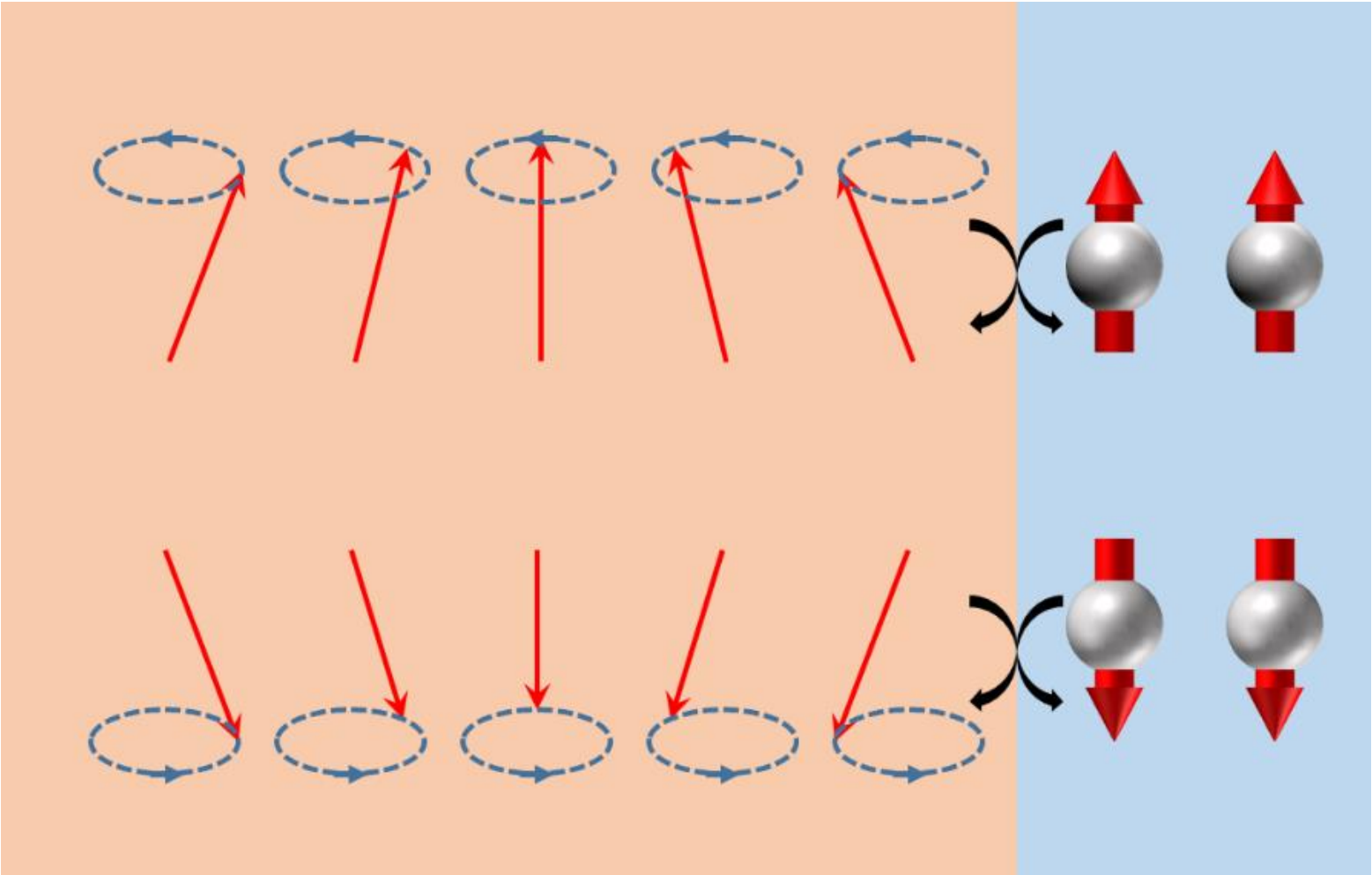

**Figure 7**. Interconversion between magnon spin current and charge spin current.

# 3. Materials for spin-orbit torques

## a. SOTs from metals and metallic alloys

The development of SOTs as a viable practical approach for electrically manipulating magnetizations has its origin in the investigation of spin Hall effects in metallic systems [13], [75]–[77]. Seminal starting points for metallic systems were the electrical detection of spin Hall effects in non-local devices [78], as well as via spin pumping [79]. Even though these early experiments suggested that spin Hall effects could be very small, soon thereafter it was shown that current-induced torques are sufficiently large to manipulate and switch magnetizations in a ferromagnetic layer adjacent to certain metallic layers, such as platinum, tantalum, and tungsten [7], [8], [80]. In a simple model, neglecting spin-memory loss and non-perfect transmission at interfaces, the normalized SOT efficiencies $\theta_{SOT}$ for damping-like torques associated with spin accumulations with polarizations in the interfacial plane and perpendicular to the applied current, are directly given by the bulk SOT efficiency [81], [82]. However, in general, it is also possible that interfacial effects contribute to the SOTs [41], [42], [83], and thus the interpretation of the experimentally measured SOT efficiencies may be more complex [81], [84]. Furthermore, interface quality and microstructure may influence the measured values. For simplicity, we refer to experimentally deduced spin Hall angles interchangeably as damping-like SOT efficiencies, even if they have not been directly obtained from measurements of electric current induced magnetic torques, *e.g.*, in the case of spin-pumping inverse spin Hall effect measurements [85], [86].

While measurements of the spin-torque efficiencies, $\theta_{SOT}$, for platinum have varied considerably in the literature, commonly accepted values are around $\theta_{SOT} \approx 0.10$ for Pt layers that are a few nm thick [32], [37], [33], [87]–[91]. Similarly, experimentally observed SOT efficiencies for Ta are $\theta_{SOT} \approx -0.10$ [37], [7], [92], and for W $\theta_{SOT} \approx -0.33$ [80]. Note that for W the large SOT efficiency is associated with β-W, which typically is only stabilized for thin films of a few nanometers thickness and the structure transforms to α-W for thicker films [80], [93]. In these materials, it is possible that the SOT efficiencies are dominated by intrinsic spin Hall effects, for which the transverse spin Hall conductivity is associated with the Berry curvature of the electronic bandstructure [94], [95]. Alternatively, large spin Hall effects may also be achieved via including impurities for strong spin-orbit coupling effects,[96], [97]. However, for alloys, enhancements of the spin-torque efficiencies may also originate from increased resistivities, since the SOT efficiency is given by the ratio of the spin Hall conductivity to the charge conductivity. Damping-like SOT efficiencies have been measured for the following alloys: $\theta_{SOT} \approx -0.24$ for $Cu_{99.7}Bi_{0.03}$ [98], $\theta_{SOT} \approx -0.13$ for $Cu_{99.5}Pb_{0.05}$[99], $\theta_{SOT} \approx 0.04$ for $Cu_{80}Pt_{20}$ [100], $\theta_{SOT} \approx 0.10$ for $Au_{93}W_7$[101], $\theta_{SOT} \approx 0.5$ for $Au_{90}Ta_{10}$[102], $\theta_{SOT} \approx 0.14$ for $Pt_{80}Al_{20}$ [103], $\theta_{SOT} \approx 0.16$ for $Pt_{85}Hf_{15}$ [103], $\theta_{SOT} \approx -0.2$ for $W_{70}Hf_{30}$ [104], and $\theta_{SOT} \approx 0.35$ for $Au_{25}Pt_{75}$ [105]. Interestingly, even repeated changes of sign of the SOTs have been observed for Au-Cu alloys [106]. In general, the exploration of different metallic alloys and compounds is still a vast open opportunity, and theoretical methods may help to identify some of the promising materials candidates, such as other A15 compounds, similar to β-W [107], [108].

As already mentioned, the interfacial microstructure may play an important role in determining the overall effective SOT efficiency. Towards this end, several studies have indicated that the presence of oxygen can change SOTs significantly [109]–[112]. In particular, oxidation of W can lead to $\theta_{\mathrm{SOT}} \approx$ -0.49 [110], while oxidation of Pt can result in even more remarkable enhancements up to $\theta_{\mathrm{SOT}} \approx 0.9$ [111], [113]. Additionally, it has been shown that the oxidation of Co at the Co/Pt interface can enhance the SOT efficiency to $\theta_{\mathrm{SOT}} \approx 0.32$ [112]. Even though the oxygen-induced SOT sign-reversal in Pt/Co bilayers [109] is attributed to the significant reduction of the majority-spin orbital moment accumulation on the interfacial nonmagnetic layer [49], the exact mechanism for the enhanced SOTs remains unclear at this point. It is possible that the enhancement is related to increased resistance, similar to the alloys discussed above. In fact, multilayer samples, which gives rise to additional interfacial scattering, can also result in enhanced SOTs. For Pt/Hf multilayers a SOT efficiency of $\theta_{\mathrm{SOT}} \approx 0.37$ [114] was measured and for Pt/Ti multilayers $\theta_{\mathrm{SOT}} \approx 0.35$ was observed [115].

Another class of metallic materials of recent interest with respect to large SOTs is that of semimetals, in particular, Weyl semimetals [116]. An important aspect of Weyl semimetals is that broken symmetries (such as time-reversal or inversion symmetries) give rise to so-called Weyl nodes in the electronic bandstructure, which are topologically required to exist and are important for transport when they are close in energy to the Fermi energy. At the same time, these Weyl nodes are sources of Berry flux, and the associated Berry curvatures are important for possibly strong intrinsic spin Hall effects. Simultaneously, symmetry considerations are also fundamentally important for spin current generation [117]. The reduced symmetries associated with Weyl semimetals may therefore also enable the generation of SOTs with novel symmetries beyond the common ones, which are associated with spin accumulations polarized within the interfacial plane and perpendicular to applied electric fields. Indeed ab-initio calculations suggest that spin Hall effects in $MoTe_2$ and $WTe_2$ can be highly anisotropic and, based on experimental charge conductivities, bulk materials may have large SOT efficiencies with maximum values for $MoTe_2$ of $\theta_{\mathrm{SOT}} \approx$ -0.72 and $WTe_2$ of $\theta_{\mathrm{SOT}} \approx$ -0.54 [118]. Experimentally, large SOTs with multiple components have been reported for few monolayers of $MoTe_2$, even though such thin films are no longer Weyl semimetals. They exhibit a complex behavior depending on layer thickness [119], and the resultant SOTs are large, $\theta_{\mathrm{SOT}} > 0.2$, for both in-plane and out-of-plane components [120]. Recently, SOT induced magnetization switching in $MoTe_2/Ne_{80}Fe_{20}$ heterostructures was demonstrated [121]. At the same time, experiments with monolayer $WTe_2$ indicate that for currents applied along certain crystalline orientations, there can be pronounced SOTs associated with spin accumulations that are polarized perpendicular to the interface plane [39]. Conversely, measurements in thicker exfoliated $WTe_2$ are sufficiently large, $\theta_{\mathrm{SOT}} \approx 0.5$, for successful electrical switching of magnetizations [122], and more recently even larger SOTs, $\theta_{\mathrm{SOT}} \approx 0.8$, were reported for sputtered $WTe_2$ films [123] enabling magnetization switching at power densities at or below those that can be achieved with common heavy metals.

## b. SOTs from topological insulators

Spin-polarized surface states in topological insulators are able to generate SOT on an adjacent magnetization similar to the spin Hall effect and the Rashba-Edelstein effect. Because topological insulators are, in principle, surface conductive but insulating in the bulk, they could potentially provide much larger SOT for the unit applied current than heavy metals due to the spin momentum locking effect. In this regard, topological insulators can serve as promising candidates for magnetic memory devices with lower power consumption. SOT from topological insulators was first experimentally demonstrated simultaneously in $(Bi_xSb_{1-x})_2Te_3$/Cr-$(Bi_xSb_{1-x})_2Te_3$ bilayer structure by second harmonic measurements [124] and in a $Bi_2Se_3/Ni_{80}Fe_{20}$ heterostructure in 2014 using spin-torque ferromagnetic resonance [73]. In the first case, a SOT efficiency $\theta_{\mathrm{SOT}}$ was measured to be as large as 425 at a cryogenic temperature (1.9 K) [124], while for the latter, a SOT efficiency of $\theta_{\mathrm{SOT}} = 2.0$ to 3.5 was quantitatively measured. Those numbers are more than one order of magnitude larger than that $\theta_{\mathrm{SOT}}$ in conventional heavy metal systems. In the former case, current-driven magnetization switching by topological insulators was also realized in this system, with a switching current density $J_{\mathrm{sw}} \approx 8.9 \times 10^4\ \mathrm{A/cm^2}$, which is more than two orders of magnitudes lower than conventional heavy metal systems.

In 2014, two independent groups have also reported the inverse process of SOT using the spin pumping technique [125], [126].

Soon after, SOT from topological insulators was demonstrated in various topological insulator/ferromagnet systems and through different measurement approaches, *e.g.*, spin-pumping, loop shift measurement and second harmonic measurement (Fig. 8a) [22], [74], [127]–[133]. Among the reported values of $\theta_{\rm SOT}$, a large discrepancy exists, where $\theta_{\rm SOT}$ can range from 0.001 to 425. The underlying physics for such discrepancy is still inconclusive, but it generally can be attributed to the variation of topological insulator quality, different measurement temperatures, and interfacial spin transparency. By depositing a ferromagnetic or ferrimagnetic layer with adequate perpendicular anisotropy on the top of topological insulators, magnetization switching driven by the SOT from topological insulators was achieved at room temperature with low switching current densities of $J_{sw} \approx 10^5\ \mathrm{A/cm^2}$ (Fig. 8a) [128], [129], [132]–[136]. Among different TIs, $(Bi_xSb_{1-x})_2Te_3$ generally possesses a larger $\theta_{\rm SOT}$ than $Bi_2Se_3$, offering a lower switching current density due to being more insulating in the bulk [133]. Sputtered polycrystalline $Bi_2Se_3$ was also shown to yield a large $\theta_{\rm SOT}$ and switching was also realized with a low $J_{sw}$ [135], [137]. The use of sputtered thin film growth may be a practical approach to be adopted in current semiconductor manufacturing. Future directions for implementing SOT for magnetic memory devices include continuously improving $\theta_{\rm SOT}$ and reducing $J_{sw}$ for better efficiency, which can be potentially achieved by improving film quality and enhancing interfacial spin transparency.

For SOT applications, besides the $J_{sw}$ which can be significantly reduced with large $\theta_{\rm SOT}$, the writing power density $P_{\rm W} = J_{sw}^2\rho$ from Ohmic loss is another crucial factor for consideration. We summarize the resistivity and the normalized power consumption [proportional to $\rho/(\theta_{\rm SOT})^2$] in a series of representative material systems: topological insulators [20], [124], [128]–[131], [133]–[136], [138], heavy metals [7], [32], [80], [139], 2D materials $WTe_2$ [122], $WTe_x$ [140], [141], $MoTe_2$[121], and $PtTe_2$[142], Rashba interfaces including STO/LAO [143] and Bi/Ag [144], and antiferromagnets [145], [146] (Fig. 8). Note that we do not compare $J_{sw}$ here since it depends on many extrinsic parameters. We observe that topological insulators have much higher SOT efficiency in general, and also a much lower power consumption even by considering the higher resistivity. Nevertheless, improving the conductivity of topological insulators will be a direction to further reduce the power consumption of SOT devices in the future. Likewise, device reliability, compatibility with current manufacturing technology, and yield in mass production also need to be considered.

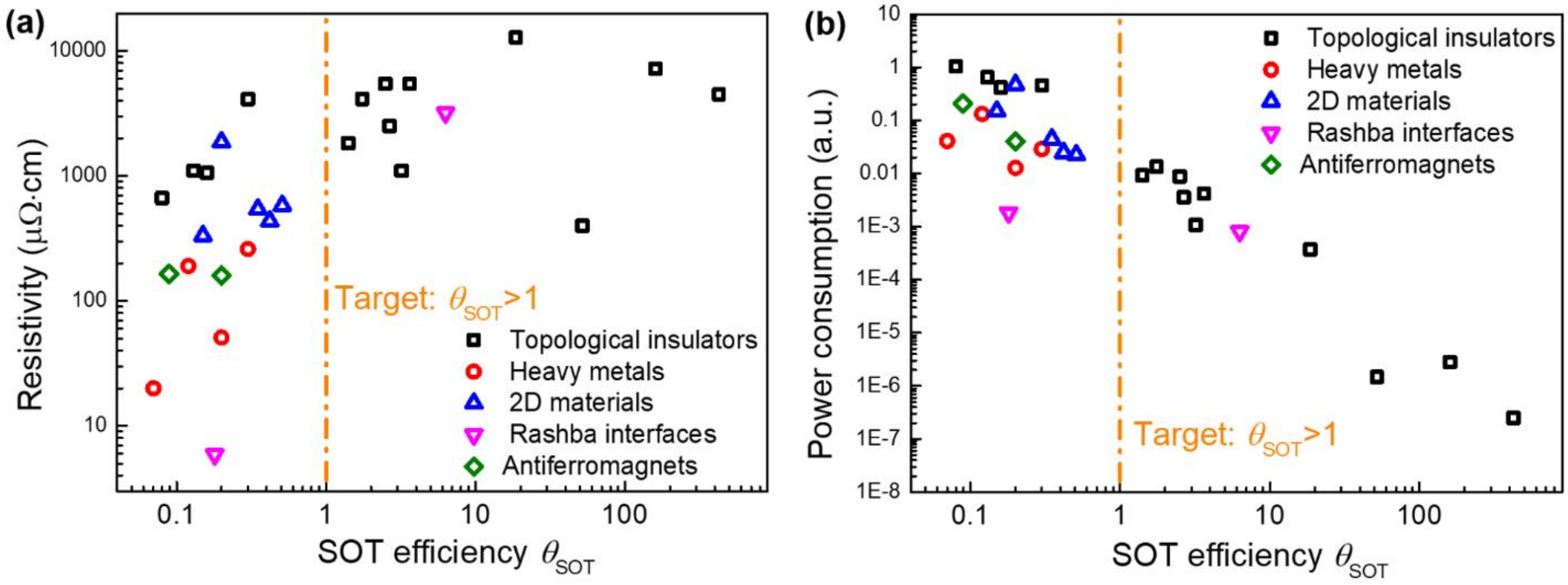

**Figure 8**. A summary of the resistivity (a) and the normalized power consumption (b) as a function of SOT efficiencies for a series of representative material systems. The target of $\theta_{\rm SOT}$ is beyond 1 for practical MRAM applications (see Fig. 1).

## c. SOTs from 2D materials

For SOT-MRAM, two improvements are required for practical applications. One is reducing switching current density, and the other is eliminating the need for a bias field. Two-dimensional (2D) van der Waals materials have shown potential in providing solutions for these two challenges [39], [119], [122], [147]–[151].

Besides van der Waals topological insulators like $Bi_2Se_3$ family, other 2D materials, some of which are topological metals [152], could also potentially provide giant SOT efficiency thanks to spin-momentum locking of topological surface states [153]. Similarly, topological metals such as Weyl semimetals are also predicted to have a strong spin Hall effect [154] as recently verified by experiments [39], [122], [150].

So far, SOT switching has only been achieved through current-induced in-plane spin polarization in nonmagnetic material/magnet heterostructures, which is not as energy-efficient as SOT switching using current-induced out-of-plane spin polarization [39]. It requires an external field to achieve deterministic switching for perpendicularly magnetized samples. 2D materials could provide unconventional SOTs that potentially allow for field-free energy-efficient switching of perpendicular magnetizations. MacNeill et al. demonstrated that an out-of-plane spin polarization could be induced in $WTe_2$ due to its reduced crystal symmetry ($C_{2v}$) when the current is applied along the low-symmetry *a*-axis (Fig. 9) [39]. However, direct demonstration of field-free switching of perpendicular magnetization using out-of-plane spin polarization from low symmetry crystals remains elusive.

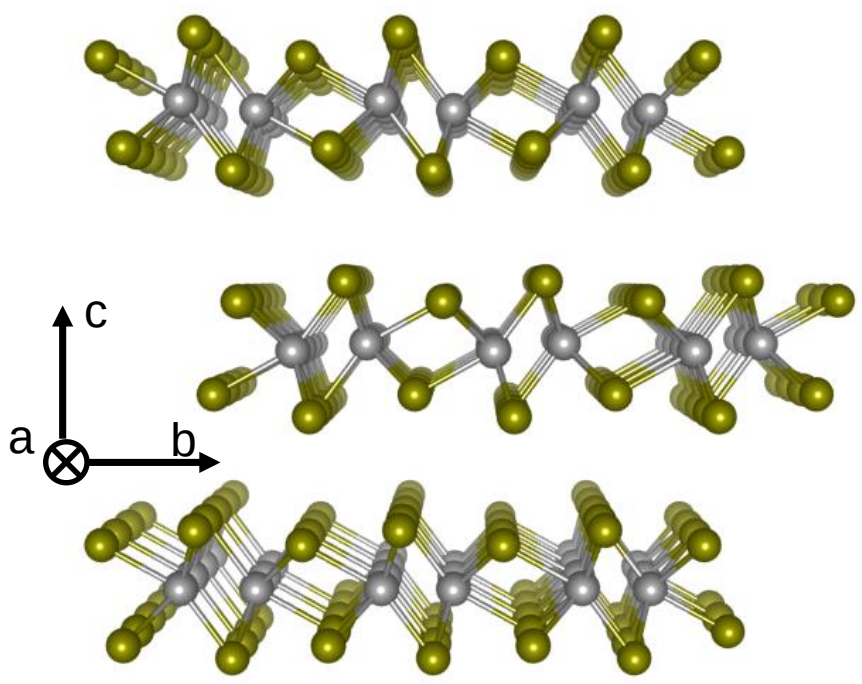

**Figure 9**. Crystal structure of $1T_d$-$WTe_2$, where the grey and sycamore colored atoms represent W and Te, respectively. Note that there is only one mirror plane – *bc* plane.

For SOT applications, the critical requirement is integrating large-scale single-crystal 2D materials with complementary metal oxide semiconductor (CMOS) technology [151]. However, spintronic properties of large-scale 2D materials are rarely explored. For conventional SOT generation, single-crystal quality may not be critical. Initial studies on monolayer Transition metal dichalcogenides grown by chemical vapor deposition (CVD) [149] and topological insulators deposited by magnetron sputtering [135], [137], [155] have shown that polycrystalline and amorphous 2D (van der Waals) materials can generate large SOTs. Single-crystal quality is critical for generating out-of-plane uniform spin polarization since different crystalline domains, and different film thicknesses could have out-of-plane spin polarization with opposite signs, which makes SOT switching non-deterministic. One potential method for integrating large-scale single-crystal 2D materials with CMOS technology could be a two-step growth: first, we achieve direct growth of single-crystal 2D material using molecular beam epitaxy, CVD, or other epitaxy growth methods on crystalline substrates; second, we transfer 2D material to the CMOS-compatible substrates, which are amorphous or polycrystalline [156].

## d. SOTs with oxides and magnetic insulators

Recent advances in film growth and synthesis techniques have enabled the growth of high-quality oxide heterostructures with very smooth interfaces. This allows for engineering novel electronic properties at the oxide interfaces [157]. One of the central oxide material systems is the interface between two wide-band-gap insulators, $SrTiO_3$ (STO) and $LaAlO_3$ (LAO), as a conductive 2DEG is formed at the interface and possesses exotic properties including superconductivity [158] and magnetism [159]. Furthermore, due to the broken interfacial inversion symmetry, the 2DEG confined in the vicinity of polar (LAO)/non-polar (STO) interface experiences a strong electric field directed perpendicular to the conduction plane [160]. Consequently, the LAO/STO interface possesses a strong Rashba spin-orbit coupling that leads to a strong coupling between the orbital and spin degrees of freedom.

The presence of strong Rashba spin-orbit coupling in the *d*-bands of the 2DEG at the STO/LAO interface has been reported in various earlier studies [161]–[163]. In addition, the Rashba spin-orbit coupling allows for the generation of spin accumulation in the LAO/STO interface when a charge current flows in the 2DEG [14], [164], which is useful for spintronic applications. Experimentally, an extremely strong charge current-induced Rashba field in the 2DEG layer was reported [165], which verifies the presence of strong Rashba spin-orbit coupling predicted at the STO/LAO interface. More recently, the inverse Edelstein effect at the STO/LAO interface has been demonstrated by using the spin pumping technique [166]–[168], which shows a notable spin-to-charge conversion. Furthermore, a long spin diffusion length over 300 nm in the 2DEG channel has also been reported [169], [170]. The above observations suggest that the STO/LAO heterostructure can be a good spin detector as well as spin channel, which is promising to advance the realization of long sought spin transistors.

The direct charge-to-spin conversion at the STO/LAO interface is an essential step for implementation in modern magnetic memory using the spin-orbit torque driven magnetization switching scheme. Using the spin-torque ferromagnetic resonance technique, a giant room temperature charge-to-spin conversion efficiency (i.e., SOT efficiency) of 6.3 was observed in the STO/LAO/CoFeB structure [143], which is two orders of magnitude larger than the SOT efficiencies in heavy metals, such as Pt and Ta [7], [32], [91]. From the temperature-dependent spin-torque ferromagnetic resonance measurements, it is understood that the mechanism of inelastic tunneling via localized states in the LAO band gap accounts for the spin transmission through the LAO layer [143].

Besides the highly efficient charge-to-spin conversion from 2DEG in oxides, 5*d*-electron transition metal oxides can generate a large bulk spin Hall effect. It was predicted that large intrinsic spin Hall effect arises in strontium iridate ($SrIrO_3$), thanks to the spin-Berry curvature from the nearly degenerate electronic spectra surrounding the nodal line [171]. The calculated spin Hall conductivity is $2 \times 10^4\ \Omega^{-1}m^{-1}$, which is comparable to that of Pt. Several individual experiments confirmed the theory, showing a SOT efficiency ($\theta_{SOT}$) between 0.4 to 1.1 [172]–[174]. The experiment demonstrated by Nan *et al.* also showed that $\theta_{SOT}$ could be modified by epitaxially tailoring the anisotropic $SrIrO_3$ crystalline symmetry [172]. A state-of-the-art structure combining both $SrIrO_3$ and a magnetic oxide $SrRuO_3$ was developed by Liu *et al.* by a pulsed laser deposition system [175]. The epitaxy in the $SrIrO_3/SrRuO_3$ heterostructure offers a new way to control the magnetocrystalline anisotropy of $SrRuO_3$. It enabled magnetic-field-free switching of the device.

Besides working as the spin-orbit coupling layer, oxides that are magnetic insulators like ferrites can also work as efficient magnetic layers. Ferrites form a large family, including spinels, hexagonal ferrites, and garnet ferrites, having the general chemical formula of $M(Fe_xO_y)$, where M represents non-iron metallic elements. Magnetic insulators have several advantages over magnetic metals for SOT device applications. First, in a heavy metal (HM)/magnetic insulator (MI) heterostructure, the charge current only flows in the heavy metal layer, but not in the magnetic insulator layer. In contrast, in an HM/ferromagnetic metal (FM) structure, the charge current also flows in the ferromagnet, resulting in certain parasitic effects. When the heavy metal layer is replaced by a topological insulator (TI) with high resistivity, the advantage of zero shunting currents in the magnetic insulator film becomes particularly important [128]. Moreover, interfacing a topological insulator with a conductive ferromagnet can result in a significant modification or even complete suppression of the topological surface states in the topological insulator layer [176], [177]. The use of a magnetic insulator can effectively avoid the shunting current; topological surface states in a TI/MI structure can also be well preserved. Avci *et al.* and Li *et al.* demonstrated the SOT switching experiments and measured large damping-like torques, in Pt/$Tm_3Fe_5O_{12}$ [178] and Pt/$BaFe_{12}O_{19}$ [179], respectively. Shao et al. demonstrated that the SOT increases as the interface magnetic density increases in W/$Tm_3Fe_5O_{12}$ [180], which confirms the theoretical prediction [181]. Li et al. also measured record-high SOT in a TI/MI heterostructure, showing the benefit of using magnetic insulators to preserve topological surface states [182]. For magnetic insulators, the intrinsic Gilbert damping constant is usually much lower than magnetic metals; this is significant for SOT oscillator applications, where the current threshold for self-oscillations decreases with the damping [58], [183]. Low damping also promotes a high-speed domain wall motion. In the experiments demonstrated separately by Vélez *et al.* and Avci *et al.*, the velocities of domain wall motion in a Pt/$Tm_3Fe_5O_{12}$ device can be very high with a small current threshold for domain wall-flow [184], [185].

## e. SOTs with antiferromagnets

The above discussion about metallic systems focused just on non-magnetic systems. However, the mechanisms that give rise to the SOTs also exist in magnetically ordered materials and may be influenced by additional symmetry breakings due to the magnetic order. Towards this end, antiferromagnetic metals are of large interest, since their magnetic order due to their vanishing net magnetic moment is typically robust against externally applied fields. An early experimental and systematic theoretical study of different CuAu-I-type antiferromagnets revealed that in Mn-based alloys the SOTs increase systematically for alloys that included heavier elements, with the largest spin-orbit torque efficiencies observed for epitaxial c-axis oriented PtMn with $\theta_{SOT} \approx 0.086$ [186], [187], which is comparable to what is observed for elemental Pt thin films. These measurements were performed with Cu layers in between the antiferromagnetic and ferromagnetic layers to avoid complications from direct exchange coupling. Removing this Cu layer may lead to a significant increase in the SOT efficiency, and for IrMn in direct contact with $Ni_{80}Fe_{20}$ $\theta_{SOT} \approx 0.22$ was measured [188] and $\theta_{SOT} \approx 0.24$ for PtMn in direct contact with Co [189]. Interestingly, the spin Hall effects in these alloys can be highly anisotropic with a factor of two difference for different crystalline orientations, such as *a*-axis vs. *c*-axis growth in PtMn or IrMn [187]. Subsequently, similar large anisotropies were observed for $IrMn_3$ thin films, where films with [001] oriented interfaces had SOT efficiencies of $\theta_{SOT} \approx 0.2$, which was almost twice as large as for [111] oriented films [190].

The symmetry breaking in antiferromagnets may also have more profound consequences than just generating an anisotropy for spin Hall conductivities; it may also generate profound new opportunities for coupling electric charge to spin degrees of freedom [191]. Antiferromagnetic spin structures with net chirality can give rise to anomalous Hall effects [192]–[194], and also may generate strong spin Hall effects [195]. Experimentally strong anomalous Hall effects have been observed for $Mn_3Sn$ [196] and $Mn_3Ge$ [197]. For ferromagnetic materials, it was already shown experimentally that anomalous Hall effects also give rise to concomitant spin currents, which can generate SOTs with new symmetries beyond those that are expected from ordinary spin Hall effects [198]–[200], and thus may similarly be expected in chiral antiferromagnets. The chiral antiferromagnetic spin structure gives rise to magnetic spin Hall effects in $Mn_3Sn$, which reverses sign upon magnetization reversal [201]. Related SOTs from magnetic spin Hall effects with new symmetries upon magnetization reversal have also been observed for $Mn_3Ir$ [202]. In addition, current-induced out-of-plane spin accumulation was observed on the (001) surface of the $IrMn_3$ antiferromagnet, which can be utilized for field-free SOT devices [203].

Given that the magnetic structure has a profound effect on the SOTs generated with metallic antiferromagnets, the natural question arises, whether exchange bias also has any influence on SOTs. Exchange bias refers to a unidirectional anisotropy that develops in a ferromagnetic layer, when an adjacent antiferromagnet is cooled through its Néel ordering temperature, or more precisely through its blocking temperature while being coupled to the ferromagnet [204]. Cooling with different magnetization states of the ferromagnet is expected to lead to different microscopic magnetic states for the antiferromagnet, which is then reflected in different magnitude or orientation of the uniaxial anisotropy. However, in two independent experiments using IrMn, there was no correlation observed between different exchange bias configurations and SOTs [205], [206]. Nevertheless, the exchange bias from the antiferromagnetic layers can provide additional advantages for applying SOTs. Namely, since conventional SOTs are related to spin accumulations with their polarization within the interfacial plane, they cannot be used for deterministic switching of magnetic layers with magnetizations perpendicular to the interface without any additional symmetry breaking. Such symmetry breaking can occur via applying an additional in-plane magnetic field, but this is not a very practical approach for actual device applications. Alternatively, the symmetry breaking can be provided by the uniaxial anisotropy due to exchange bias. Several research groups have demonstrated that SOTs from metallic antiferromagnets can provide deterministic switching of perpendicular magnetizations and alleviate the need for any additional externally applied magnetic fields [207]–[211]. Interestingly, experiments have shown that the exchange bias direction can be switched by using SOTs [212].

Besides the torques that metallic antiferromagnets can exert on adjacent ferromagnets, some metallic antiferromagnets can also offer a very new way of electrically manipulating magnetic structure. Namely, in antiferromagnets where the individual spin sublattices are associated with crystal lattice sites that have reciprocal local inversion symmetries, electric currents can also generate so-called Néel (or staggered) SOTs, so that the direction of the SOTs has opposite orientation for each sublattice. This was first theoretically predicted for $Mn_2Au$ [213]–[215]. Subsequently, such electrical switching behavior was observed for CuMnAs [216], [217] and $Mn_2Au$ [218], [219]. So far, the efficiencies of these torques are relatively low, and thus electric manipulation of antiferromagnet states requires very high current densities of about $10^6$ A/cm$^2$ to $10^7$ A/cm$^2$, which even then only result in a partial rearrangement of antiferromagnetic domains [220], [221]. Therefore, the role of heating and electromigration needs to be carefully evaluated when investigating electrical SOT phenomena manipulating the magnetic structures of antiferromagnets [222]–[226].

## f. SOTs with ferrimagnets

While the majority of the research employs ferromagnetic material as the free layer in the study of SOT induced magnetic dynamics, some non-conventional magnetic materials, including antiferromagnet and ferrimagnet, have attracted great interest very recently due to the rich physics and promising application prospects. It is generally believed that the high-frequency dynamics associated with the antiferromagnetic mode in those materials can lead to high-speed magnetic switching from SOT. Moreover, the zero or reduced net magnetic moment will decrease the stray magnetic field generated from each device, minimizing the magnetic cross-talk among neighboring bits. Meanwhile, the low magnetic moment also provides extra robustness against external field perturbation. The recent progress in antiferromagnetic spintronics has been summarized in a few review articles [227]–[230]. Interested readers are encouraged to refer to that literature for an in-depth understanding. In this session, we will focus on the current status and outlook on ferrimagnet-based SOT studies.

Ferrimagnets, to a certain extent, combine advantages from ferromagnets and antiferromagnets, which provides more flexibility in material and device engineering. First of all, similar to antiferromagnets, ferrimagnets have antiparallel aligned spin sublattices, which lead to the high-frequency dynamics mode as well as reduced net magnetization. In certain situations, the net moment within a ferrimagnet can be tuned to be zero by compensating the magnetization from the two sublattices through chemical composition adjustment, forming an effective antiferromagnet [231]. Meanwhile, unlike antiferromagnets, where it is usually difficult to have efficient mechanisms for probing magnetic states, ferrimagnets can exhibit magneto-transport phenomena similar to those in a regular ferromagnet, allowing for easy magnetic reading [232]. Particularly, finite magneto-transport coefficients, such as anomalous Hall resistance [233] and tunneling magnetoresistance [234], in principle, are expected to exist in ferrimagnetic samples both with and without a net moment. This is due to the fact that at the Fermi level, the spin-dependent electron density of states is usually not equal between the two spin sub-bands, leading to different contributions from the two sublattices to electrical transport effects [233]. Because of this exact reason, in theory, it is even possible to design compensated ferrimagnets with half metallicity, i.e., a magnet with zero moment but 100 % spin polarization at the Fermi level, as is proposed by van Leuken and de Groot [235].

In terms of writing methods for ferrimagnet-based spintronic devices, SOT provides a convenient mechanism, where spins injected via the spin Hall effect [7], or Rashba-Edelstein effect [8] can directly interact with the two spin sublattices and lead to magnetic dynamics constructively. In Ref. [236], Gomonay et al. discussed the spin current-induced magnetic dynamics in a system with two antiparallel aligned sublattices. In a ferrimagnet or antiferromagnet with uniaxial anisotropy, the damping like spin-orbit torque on the two sublattices has the form of $\boldsymbol{\tau_{D,i}} = J_{s,i}\boldsymbol{m_i} \times (\boldsymbol{\sigma} \times \boldsymbol{m_i})$, where $\boldsymbol{m_i}$ ($i$ = 1, 2) represents the magnetic moment orientation of the two sublattices, $J_{s,i}$ is the magnitude of the spin current acting on each sublattice, and $\boldsymbol{\sigma}$ is the injected spin orientation that accounts for SOT. Under the assumption of a strong exchange coupling $\boldsymbol{m_1} = -\boldsymbol{m_2}$, the two spin torques have the same sign, which constructively rotates the magnetic ordering in a ferrimagnet and leads to magnetic switching [237]–[239]. The SOT induced switching in rare earth-transition metal alloy ferrimagnets was experimentally verified in 2016 [240], [241], where it was shown that the spin current from heavy metal could cause magnetic switching in alloys made from Co, Fe, Gd, Tb, etc even when the net magnetic moment is very close to zero [242]–[246]. More quantitatively, it was demonstrated that the effective field from the damping like torque diverges at the magnetic moment compensation point, following the relationship of the damping-like effective field, $H_{DL} \propto \frac{1}{M_s}$, a result that is expected from the SOT picture described above. Here, $M_s$ is the net saturation magnetization. It should be noted that the divergent behavior of $H_{DL}$ close to the compensation point does not imply that the threshold current can be reduced to zero. In reality, for real applications, the requirement of a decent thermal stability requires that the coercive field needs to be very large in a nearly compensated ferrimagnet to keep the thermal barrier $E = M_s H_c$ finite [240]. Therefore, the increase of $H_{DL}$ will be compensated with the corresponding enlargement of $H_c$. Besides employing heavy metals to switch a ferrimagnet, studies have been carried out where topological insulators [128], [130] are used as the source for SOT. Higher efficiency over traditional heavy metals are demonstrated in these experiments, due to the higher effective SOT efficiency.

Besides rare earth-transition metals, quite a large number of magnetic insulators belong to the family of ferrimagnets, due to the oxygen atom mediated super-exchange. SOT induced magnetic switching has been demonstrated in these materials [178], [180], [182], which shows comparable or even lower threshold current compared with magnetic metals. Insulators, in general, are believed to be more favorable candidate materials for low power spintronic applications, due to the absence of Joule heating. One question that was unclear was whether efficient spin injection can happen at the interface between a magnetic insulator and a spin Hall metal or topological insulator. In these recent experiments, a large spin mixing conductance was demonstrated, which ensures efficient SOT upon those insulating ferrimagnets.

Heusler alloys are another important category of magnetic materials, many of which exhibit ferrimagnetic ordering. For example, Heusler alloys with Mn elements usually have antiparallelly aligned sublattices and have reduced saturation magnetization. Interested readers are suggested to refer to review articles of refs. [234] and [247] for a detailed discussion on the general properties of these materials. In terms of SOT induced magnetic dynamics, it has been shown that magnetic switching or oscillation can be driven by spin Hall effect from adjacent heavy metal layers or even by the SOT within the Heusler alloys themselves [248]–[250], where the latter one is due to the breaking of symmetry in specific Heusler alloys.

One of the main motivations for the employment of materials with antiferromagnetic coupling is to obtain a faster operation speed. Recently, the speed advantages associated with ferrimagnets are demonstrated in multi-domain ferrimagnets in the domain wall motion regime [184], [185][251], [252]. Similar to the previously reported synthetic antiferromagnetic case [253], the domain wall speed largely increases when the angular momenta between the two sublattices are roughly canceled out, due to the mutual interactions between SOT, Dzyaloshinskii–Moriya interaction (DMI), and inter-lattice exchange coupling [251]. Particularly, experimentally it is confirmed that the absence of Walker breakdown keeps domain wall speed from saturating as the driving current increases [251], [252]. Currently, a record high domain wall speed of a few kilometers per second has been reported in SOT driven experiments [185], [252], [254], as is mentioned in previous sections.

For the goals that are expected to be reached in the near future, one important topic on ferrimagnetic SOT study is to realize integrated MTJ devices that can be switched by SOT and be detected by tunneling magnetoresistance. As is discussed above, compared with antiferromagnets, ferrimagnets have finite spin polarization at the Fermi surface, allowing for large magnetoresistance ratio. Experimentally, the integration of SOT based ferrimagnet devices with an MTJ structure still faces several material challenges. For example, the active chemical properties of rare earth elements in their metallic state usually determine that they do not form a sharp interface when being in contact with an oxide tunneling barrier such as MgO [255]. Heusler alloy ferrimagnets are believed to be a promising candidate for achieving high tunneling magnetoresistance [234], [247]. So far, despite the theoretical prediction [235][256] and preliminary experimental evidence [257], [258] on the existence of half metallicity in certain Heusler alloys, demonstrations of a tunnel magnetoresistance (TMR) ratio that is larger than standard CoFeB based MTJs still remain to be scarce [259], [260]. The small bandgap in spin sub-bands and the atomic level roughness among different grains could be the origins of the diminished TMR in existing experiments [258], [261]. For ferrimagnetic insulators, achieving a high ON/OFF ratio through the mechanisms of Hall resistance or (spin) Hall magnetoresistance could be one of the key pre-requisites for practical applications of these materials.

Besides improving the efficiency for magnetic reading in SOT-based ferrimagnet devices, increasing the SOT efficiency and further reducing the switching current will be another aspect for future study. As is discussed earlier, standard SOT model determines that the effective field scales with the net magnetic moment in a manner of $H_{\mathrm{DL}} \propto \frac{1}{M_s}$. However, experimental evidence exists that $H_{\mathrm{DL}}$ becomes divergent at a rate faster than $\frac{1}{M_s}$, providing the potential for achieving a higher switching efficiency [245]. While the exact mechanism remains to be further studied, it is believed that the internal interactions among the two sub-lattices, as well as the coherence of injected spins across the different magnetic layers, account for these behaviors [262].

Finally, demonstrating robust sub-nanosecond magnetic switching or ultra-high frequency magnetic oscillation could be another milestone that will be reached within the next few years. As is mentioned above, the ultrafast magnetic domain wall motion has been observed in multi-domain ferrimagnets. Extrapolating this high speed magnetic switching into the single domain regime will be highly useful for the realization of standard single bit SOT-MRAM cells. While sub-nanosecond switching has been reported in various circumstances in the study of SOT and STT spintronics, it remains to be demonstrated that ultra-low writing bit error rate can be achieved in those devices [263]–[265]. Particularly, a writing bit error rate lower than $10^{-11}$, i.e., one error for every $10^{11}$ switching events, is believed to be required for a general-purpose magnetic memory technology [266]. Ferrimagnets, with their ultrahigh intrinsic magnetic dynamic frequency, can potentially provide unique advantages along this direction. Besides magnetic switching, electrically induced high-frequency magnetic oscillation is another direction in which a ferrimagnet could play an important role. Besides the existence of high intrinsic magnetic resonance frequency, another important requirement for the excitation of sustained magnetic oscillation is low magnetic damping. It is known that the threshold current for magnetic oscillation has the dependence of $I_c \propto \alpha f$, where $f$ is the oscillation frequency. Therefore, ultra-high currents need to be applied for exciting terahertz frequency oscillations. Ferrimagnets, particularly those with a bandgap in one of the spin sub-bands, can have low magnetic damping due to the absence of spin dependent scattering of conduction electrons, which can help to reduce the threshold current [267]–[269].

## g. SOTs with low-damping ferromagnets

Loss in magnetic precession, often discussed in terms of Gilbert damping, plays a key role in essentially all types of magnetization dynamics [270], [271]. Some spintronic device applications benefit from optimally high damping, as it suppresses fluctuations in magnetic read heads [272] and undesirable switch-backs in SOT-driven perpendicular magnetic media [273]. Damping can be increased readily, for instance, by alloying or interfacing ferromagnets with some heavy elements with strong spin-orbit coupling [274]–[276].

There are also numerous applications where lower damping is desirable, particularly for lowering loss in magnetic precession. Magnetic thin-film media with low damping are therefore in demand for such devices as spin-torque nano-oscillators [277], [278] and spin-wave logic and circuits [21], [67] (including those for possible use in quantum information technologies[279], [280]). However, low-loss magnetic materials are not as straightforward to engineer. There are challenges not only in identifying materials with low *intrinsic* Gilbert damping [281], [282], but also in minimizing the total effective loss (as parameterized by resonance linewidth, Q-factor, critical current density, etc.) from *extrinsic* relaxation or linewidth-broadening mechanisms [270], [271]. These extrinsic mechanisms may include spin pumping [283], spin-memory loss [284], two-magnon scattering [285], and inhomogeneous broadening [286].

The quest for low-loss magnetic media is further complicated by practical considerations, such as the need for high magnetoresistance signals and the compatibility with commercial device fabrication processes. In this regard, ferromagnetic metals have significant advantages for practical device applications. The magnetization state of ferromagnetic metals can be detected via large magnetoresistance signals (e.g., ≈ 1% for anisotropic magnetoresistance, ≈ 10 to 100 % for giant magnetoresistance and TMR). Thin films of ferromagnetic metals can also be grown readily on Si substrates by sputtering under conditions compatible with standard device fabrication processes (e.g., growth temperature below ≈ 300 °C).

Although ferromagnetic metals are often thought to exhibit much higher damping than magnetic insulators, recent experiments point to ultralow intrinsic Gilbert damping in Heusler compounds[267], [287], [288] and CoFe alloys [289], [290]. Moreover, contrary to popular belief, most magnetic insulator (ferrite) films actually exhibit higher effective loss than ferromagnetic metals. FMR linewidths evidence this for typical ferrite films (e.g., ≈ 10 to 100 mT at ≈ 10 GHz) [291], [292] that are an order of magnitude greater than those of many ferromagnetic metals. The few thin-film insulators (e.g., YIG [293]–[296] and coherently strained spinel ferrites [297], [298]) that do show substantially lower loss are generally challenging to grow, e.g., requiring epitaxy on lattice-matched crystal substrates. Therefore, it is reasonable to expect that ferromagnetic metals will continue to be the main material platforms for practical SOT devices.

We identify the following challenges in engineering metallic ferromagnetic media for practical SOT devices with low-loss precessional dynamics:

*A. Understanding how film structures affect energy loss in precessional dynamics (Gilbert damping, two-magnon scattering, inhomogeneous broadening, etc.):* Ultralow intrinsic damping parameters of $\approx 5 \times 10^{-4}$ have been reported in sputter-grown polycrystalline CoFe alloy films [289]. These CoFe films appear to be more viable than Heusler compounds, for which it is difficult to achieve the correct structure and stoichiometry for theoretically-predicted ultralow-damping. In Ref. [289], the intrinsic Gilbert damping parameters are obtained with films magnetized out of the plane, the measurement geometry that eliminates two-magnon scattering contributions to the damping [299]. However, many potential applications require in-plane magnetization [67], [277], [278], in which case non-Gilbert relaxation may be substantial. A follow-up study [300] shows that certain seed layers result in a significantly lower loss (narrower resonance linewidth) even when the film is magnetized in-plane. A more detailed understanding of how the structure of the ferromagnetic metal (e.g., tuned by the seed layer) affects magnetic relaxation and broadening due to inhomogeneities will enable magnetic media with ultralow effective loss.

*B. Developing practical low-damping, low-moment ferromagnetic media:* A possible drawback of CoFe is its high saturation magnetization [301] (in fact, near the top of the Slater Pauling curve), which increases the critical current density for spin-torque excitation. In that regard, Fe alloyed with nonmagnetic V may be a viable alternative with similarly low intrinsic Gilbert damping[282], [302]. A recent experimental report confirms a low intrinsic damping parameter of ≈0.001 in BCC FeV alloy films [303]. Exploring other possible ultralow-damping ferromagnetic alloys with low moments that are also straightforward to grow is crucial for future SOT applications.

*C. Resolving the issue of increased damping in SOT device structures:* While some ferromagnetic alloys like CoFe and FeV have ultralow intrinsic damping, SOT applications require the ferromagnetic metal to be in contact with (or in proximity to, via a spacer material) a spin-orbit metal, i.e., the source of charge-to-spin conversion. The ferromagnetic metal must be thin (e.g., <<10 nm), since the critical current density is inversely proportional to the magnetic layer thickness. Fundamentally, this is because the conventional SOT is essentially an interfacial effect, in the sense that spins must be transferred across the interface between the spin-orbit metal and the ferromagnet.

The proximity of the ferromagnet to a spin-orbit metal introduces additional sources of loss. Some of these loss mechanisms, such as spin-memory loss [284] and two-magnon scattering [285], can be reduced (at least in principle) by properly engineering the interfacial quality or inserting an appropriate spacer-layer material. However, the spin-orbit metal presence inevitably leads to increased Gilbert damping from spin pumping [283], the reciprocal process of SOT. The magnitude of spin-pumping damping is inversely proportional to the magnetic thickness. There thus arises a conundrum for practical applications: the thin ferromagnet required for SOT-driven dynamics invariably exhibits high damping.

A possible route to overcome this thickness conundrum is to go beyond the "interfacial" scheme for SOTs. In particular, a viable solution may be a SOT that originates in the bulk of the ferromagnetic metal [34], [304]. SOT-driven precessional dynamics can be achieved in a thick single-layer magnet without any damping increase from spin pumping. A few recent experimental studies indeed point to the presence of such "bulk" SOTs in various room-temperature ferromagnets that are ≳10 nm thick [44], [249], [305]. An open challenge is to engineer a sufficiently strong bulk SOT to drive precessional switching or auto-oscillation in magnetic media, compatible with experimental fabrication processes and electrical detection schemes. Also, a natural question is whether a high bulk SOT efficiency – i.e., strong spin-orbit coupling – in a ferromagnet is accompanied by high Gilbert damping. Some recent experiments [303], [306], [307] indicate that the strength of spin-orbit coupling is not necessarily the primary factor governing intrinsic damping; rather, a small density of states at the Fermi level appears to be essential for low damping in ferromagnetic metals. There is thus reason to be optimistic that a strong bulk antidamping SOT and ultralow damping can be simultaneously achieved in engineered ferromagnetic metals.

# 4. Devices based on spin-orbit torques

## a. Three-terminal SOT memory

*SOT-MRAM, STT-MRAM, Cache memory*: SOT-induced magnetization switching can be used for the write method of MRAM. MRAM technology began with a magnetic-field writing scheme. However, because required current for magnetic-field-induced switching is relatively large, typically in the order of milliamperes, and tends to increase with the reduction of device size, focus of researches has shifted to STT-induced magnetization switching [63], [64] due to its scalability. After several breakthroughs including a perpendicular-easy-axis CoFeB/MgO MTJ [308], production of STT-MRAM has recently started. The current STT-MRAMs are mainly used as a replacement for embedded flash memories, and application to last-level cache memories is expected to start in the near future. Write pulse widths used in typical STT-MRAMs are on the order of several tens of nanoseconds, and can be reduced down to a few nanoseconds. However, further reduction of the write pulse width is challenging since the required write current is inversely proportional to the write pulse width. Accordingly, there is an increasing demand to develop MRAMs that can operate on a timescale of less than a few nanoseconds and SOT-induced magnetization switching is expected to be a prime candidate.

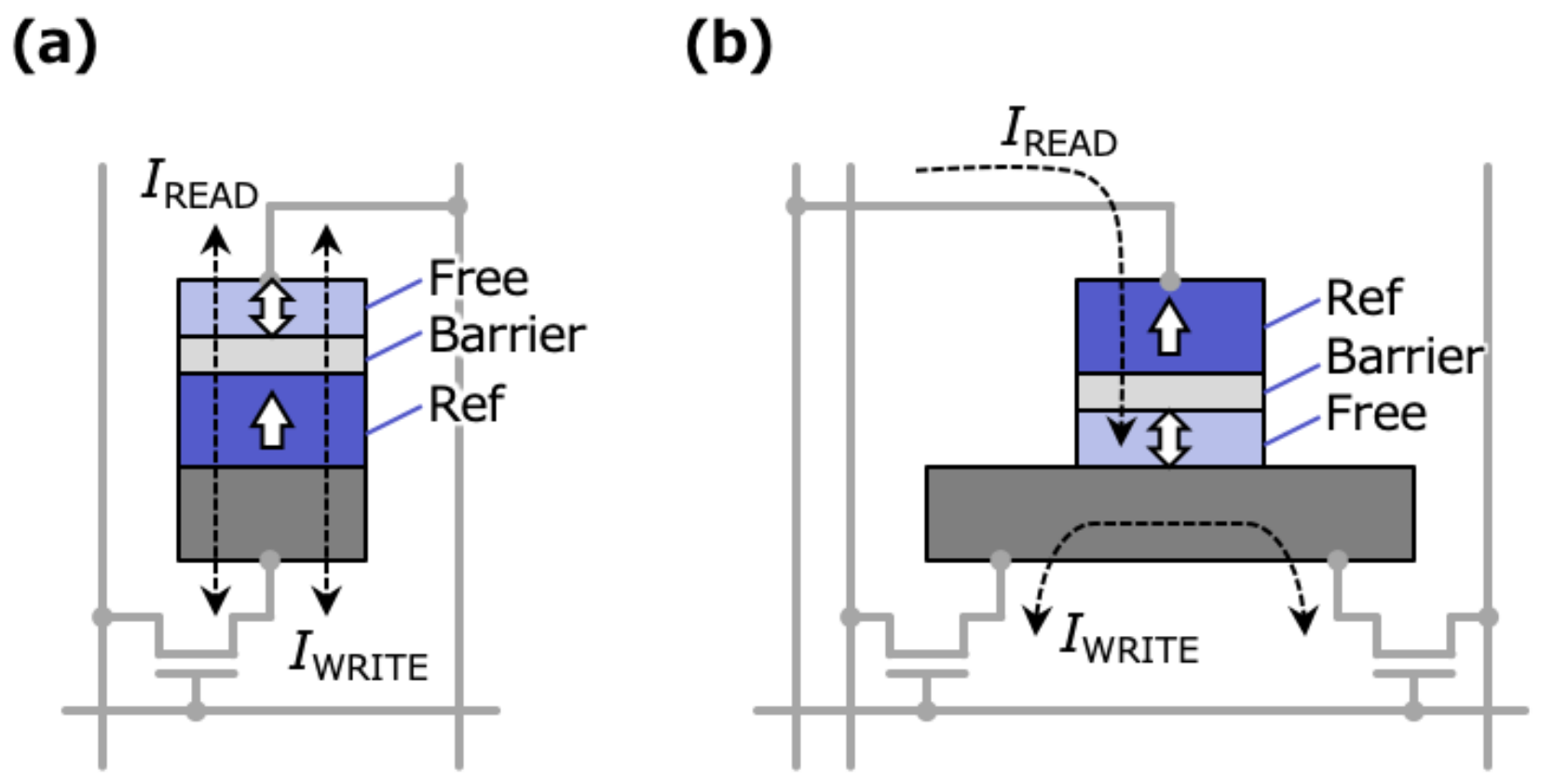

**Figure 10**. Typical cell structure of STT-MRAM (a) and SOT-MRAM (b). Exemplary magnetization directions are indicated by white arrows and write $I_{WRITE}$ and read $I_{READ}$ current paths are indicated by broken arrows.

For SOT-MRAM, a transverse current passing through the underlying heavy metal injects a spin current in the nanomagnet lying on top. This mechanism allows decoupled read and write mechanisms, unlike in STT-MRAM, leading to separate optimization of the read and write paths. The SOT mechanism can also be more efficient for magnetization reversal, leading to a lower critical current required to switch the magnet from one state to the other, thereby enabling a fast and energy-efficient write operation compared to STT-MRAM.

Typical cell structures of the STT-MRAM and SOT-MRAM are shown in Figs. 10(a) and 10(b), respectively. The STT-MRAM cell consists of a two-terminal MTJ device and one cell transistor (1T-1MTJ), whereas the SOT-MRAM cell consists of a three-terminal MTJ device and two cell transistors (2T-1MTJ). On the one hand, 2T-1MTJ cell structure inevitably requires a larger cell area than the 1T-1MTJ cell, leading to lower memory density. On the other hand, write and read current paths are different from each other in the three-terminal cell, providing the following benefits: First, the separation of the current paths provides a large operation margin for both read and write operations, allowing high-speed operation. For the two-terminal MTJ, an upper limit of the read current is determined by the current that unintentionally induces magnetization switching and an upper limit of the write current is determined by the current that unintentionally breaks the tunnel barrier. These restrictions are essentially lifted in the three-terminal cell, and short current pulses, less than a few ns, can be applied for operation [309], [310]. Second, unlike the two-terminal structure, the write current does not pass through the tunnel barrier for the three-terminal cell. This fact offers high endurance, which is a critical requirement for high-speed cache memories. Third, the cell circuit of the 2T-1MTJ is similar to the typical cell structure of static random-access memory (SRAM); the flip-flop circuit of the 6T-SRAM cell is replaced by one MTJ, and also it can fit well with the circuits of shift registers [311]. This similarity offers good compatibility with existing higher-level cache memories.

There are several structures of actual SOT-MRAM devices that are different in the direction of magnetic easy axis as shown in Fig. 11. Defining the *x* axis to be along the write current and *z* axis to be along the out-of-plane direction, each structure has the easy axis along *z* (Fig. 11(a)) [8], *y* (Fig. 11(b)) [7], *x* (Fig. 11(c)) [312], and in between the *x* and *y* (Fig. 11(d)) [313] directions. For the structure shown in Fig. 11(b), the accumulated spin in the ferromagnetic layer is collinear with magnetization and its switching mechanism is basically the same as STT switching. Thus, a magnetic field is not required for switching but the critical current increases as the pulse width decreases with an inverse proportional relation as in the case of STT switching [314]. On the other hand, the structures shown in Figs. 11(a) and 11(c) are capable of fast switching [313], [315], [316] but this requires a static external magnetic field for switching [317] because the switching is driven by electron spins orthogonal to the magnetization. Thus, elimination of the necessity of the external field is a big challenge for the application of these structures; some examples are described in section 4d. The structure shown in Fig. 11(d) was found to achieve reasonably fast switching at zero magnetic field [318].

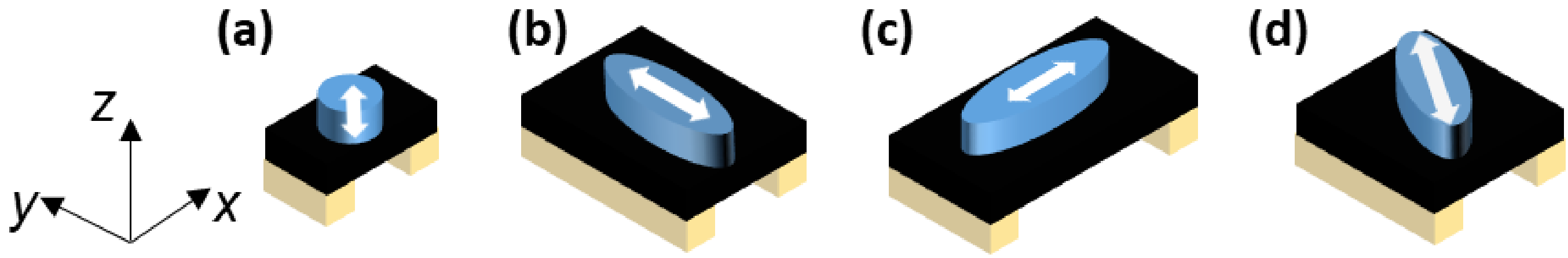

**Figure 11**. Four types of structures of SOT switching devices for memory application (drawn to scale): a) Z type with out-of-plane magnetic easy axis, an external magnetic field along the x axis is required for deterministic switching, b) Y type with an in-plane easy axis perpendicular to current direction, c) X type with an in-plane easy axis parallel to current direction, an external field along the z axis is required for deterministic switching, d) XY type with an in-plane easy axis tilted with respect to current direction. Magnetization easy axis directions of the switching layer are indicated by white arrows.

Operation of three-terminal SOT devices has been demonstrated by several groups [7], [9], [80], [312], [319]–[325]. The first demonstration was reported by Liu *et al.* using Ta/CoFeB/MgO-based MTJ. Demonstration of field-free operation of a SOT-MRAM device fabricated through CMOS-compatible 300 mm wafer process has been reported by Garello *et al*. [324] and Honjo *et al*. [325]. Garello *et al*. utilized a structure with an easy perpendicular axis (Fig. 11a);

in addition, a stray field generated by a separately prepared magnetic layer is used for field-free operation. Honjo *et al*. utilized a structure with an in-plane easy axis in the *x*-*y* plane (Fig. 11d) and high-speed operation, high thermal stability, and high thermal tolerance were reported to be satisfied simultaneously. Very recently, Natsui *et al.* demonstrated the circuit operation of 4-kB SOT-MRAM [11]. Further enhancement of device properties is expected to realize SOT-MRAMs used in integrated circuits in the future.

For practical use of SOT-MRAM, SOT-induced switching devices should satisfy several requirements, which are listed below. One of the most important criteria is that the critical current required for magnetization switching be about 100 µA or less. This is mainly because the size of the cell transistor is determined by the write current and a larger write current requires a larger cell size, eventually limiting the memory capacity. A larger write current is also undesirable in terms of lower-power operation, which is particularly critical for internet-of-things (IoT) applications. In addition, the resistance of the write current path should be small enough so that the current required for switching can be supplied with a reasonably small source voltage, typically a few 100 mV or less. Magnetization switching speed is an equally important criterion, as SOT-MRAMs are expected to be used for high-speed memories in integrated circuits that STT-MRAMs cannot access. Switching within a few nanoseconds, or ideally sub-nanosecond, is required. In terms of high-reliability operation that is an important requirement for cache memory applications, low switching error rate, and high endurance are demanded. In addition, since MRAMs attract interest due to their nonvolatility, high thermal stability is also an important factor. Furthermore, for the compatibility with CMOS processes and circuits, materials for the free layer should exhibit a high TMR ratio, and the stack structure should have a thermal tolerance up to 400 $^{o}$C, the highest temperature of standard CMOS and packaging processes.

### b. Two-terminal SOT memory

Besides the commonly employed three-terminal configuration (Fig. 12a) [7], [80], [312], two terminal SOT devices (Fig. 12b) have been developed recently for their potential capability in achieving higher storage density. Due to the nature of spin-orbit interaction, in order to generate spin accumulation at the surface of thin films for magnetic switching through the mechanism of spin Hall effect [7] or Rashba-Edelstein effect [8], one has to apply charge currents that flow within the film plane. This pre-requisite has led to the popular three-terminal design in existing SOT devices. However, the large footprint, as well as the necessity of using more than one transistor for device selection, have made this design unfavorable for applications where the bit density is one of the most important considerations, such as memory. Meanwhile, the majority of mainstream memory technologies nowadays (e.g., DRAM, flash, etc.) employ a two-terminal geometry, where the memory cell (and the backing transistor) sits at the cross point of the word line and the bit line [1], [3]. Developing SOT MRAM devices that comply with this convention can greatly simplify architecture and facilitate circuit design.

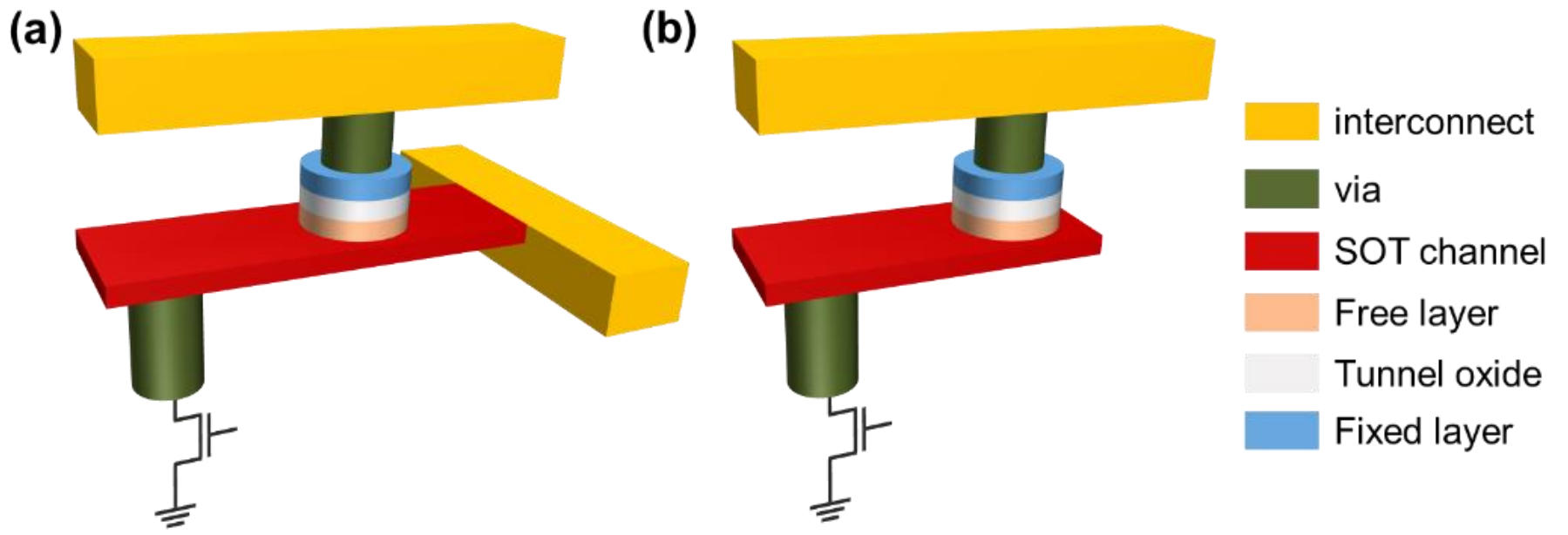

**Figure 12**. Comparison of a) three-terminal, and b) two-terminal SOT-MRAMs.

The two-terminal SOT device was first proposed in 2013 [326], where it was shown that, by tailoring the current flowing path across an MTJ, one can realize a compact design for SOT MRAM cell, fitting within the usual design principle of high density DRAM, known as '$6F^2$'. The device design in this two terminal SOT MRAM is very similar to that in conventional STT-MRAM [3] except that instead of directly sitting on top of the backing transistor, the MTJ is offset from the transistor by a small horizontal distance (Fig. 12b). The bottom layer of an MTJ and the top contact of a transistor (source or drain electrodes) are connected through a thin channel layer with strong spin-orbit interaction. During the writing event, the applied charge current will flow first vertically through the MTJ, then horizontally along the channel. Both the SOT from the underlayer and the STT from the tunneling current will exert influences onto the dynamics of the magnetic free layer, leading to magnetic switching. A. Brink et al. Ref. [327] numerically studied the combined effect of SOT and STT in magnetic switching and showed that large reductions in switching time and energy can be achieved. The concept of a two-terminal SOT MRAM cell was experimentally verified recently by Sato et al. [328] and Wang et al. [329], independently, where STT and SOT are employed jointly to switch magnetic free layers with perpendicular anisotropy. It was shown that due to the high SOT efficiency in those experiments, SOT plays the dominant role as the switching mechanism. Meanwhile, the incorporation of STT mechanism brings in an additional advantage, i.e., making the SOT-dominated switching field-free [329]. As is discussed earlier in this perspective article, realizing field-free switching of SOT MRAM cell with perpendicular anisotropy is one of the important research directions [207], [208], [330], [331]. Since STT induced switching in standard MTJs is already deterministic, adjusting the relative ratio between the two spin torques in the two-terminal device can lead to the optimal region where the advantages of low switching current and being field-free can be achieved simultaneously.

To predict the future research directions on two-terminal SOT devices, one important topic is to incorporate the newly discovered high efficiency SOT materials into the two terminal design. As is discussed in this perspective article, groups of materials with large effective SOT efficiency including topological insulators [73], [124], [128], [136], semimetals [39], [122], 2D materials [332], [333] and 2D electron gas systems [143], [166], [334] have been identified recently. In principle, the switching current density can be significantly reduced via the use of these novel materials, which will minimize the needed tunneling current across the MTJ barrier and enlarge the gap between the writing voltage and breakdown voltage. One has to note that as the growth of high quality MTJs usually requires certain seeding layers or specific crystalline textures [308], [335], [336], showing the compatibility of these novel materials with MTJ growth will be a critical step.

Besides further increasing the SOT efficiency with the usage of new materials and new mechanisms, realizing SOT induced real 'anti-damping' switching will be another important topic to visit in the study of two-terminal SOT devices. In most of the existing studies where SOT is employed to flip a perpendicularly magnetized free layer, the switching is known to be inefficient as the injected spins are orthogonal to the equilibrium orientation of magnetic moments. In this case, the threshold current is roughly proportional to the anisotropy field in the single domain configuration [317], $J_c \propto H_a$. Therefore, under the requirement of high thermal stability ($E \propto H_a$) for small MRAM cells, the needed switching current density is expected to be extremely large. As is discussed in this perspective article, a few recent experiments suggest that by harnessing the symmetry of certain crystals [39] or utilizing the magnetic moment induced spin precession [43], one can generate spins at film surface which are orientated out of plane. This will finally enable 'anti-damping' switching from SOT, where the threshold current is used to counterbalance the relatively smaller damping torque where $J_c \propto \alpha H_a$, instead of the torque from large anisotropy field [7]. Here α is the Gilbert damping coefficient. Since $\alpha$ is usually smaller than 0.01, a reduction of more than 100 times can be expected in threshold current if the switching happens in the 'anti-damping' regime.

Overall, two terminal SOT devices combine the advantages of high density of traditional MRAM cells and high efficiency of SOT, providing a competitive solution for dense, low power, and non-volatile memory. The future development of this technology will integrate new SOT materials with MTJs, achieving effective SOT efficiency larger than one while maintaining the high tunneling magnetoresistance and strong magnetic anisotropy. Meanwhile, materials and mechanisms which lead to out-of-plane spin generation will give another large boost in the switching efficiency of two-terminal SOT devices.

## c. SOT neuromorphic devices and circuits

Beside memory applications, spintronic devices can be used for in-memory and neuromorphic computing [337], [338]. It is interesting to note that standard three terminal structures have a huge potential to be used in analog in-memory computing deep neural network applications (Figs. 13a and 13b) [339], [340]. Indeed, deep neural network inference requires a massive amount of matrix-vector multiplications, which can be computed energy efficiently on memory arrays in an analog fashion. This approach, however, requires highly resistive memory device levels (greater than megaohm) with low cell-to-cell and time variability of device resistance and reading window to implement deep neural network weight memories. MRAM technology is among the most promising candidates fulfilling these requirements. However, STT cannot be used to switch high resistance MTJs. Here, one can exploit the 3-terminal geometry of SOT, where writing is not limited by the MTJ resistance level, as write and read paths are decoupled. Preliminary experimental and design works show [340] that SOT-MRAM is indeed an excellent candidate to implement ternary weights for inference accelerators running quantized deep neural networks, with MTJ resistance that can be tuned from 1 MΩ to >50 MΩ (Fig. 13c) with tight variability and no compromise on the write process. Design technology co-optimization study, where convolutional neural networks (for the MNIST [341] and CIFAR100 [342] databases) are trained and tested including resistance variation and the effect of wire inference read, shows that SOT-MRAM meets the derived specifications (Fig. 13d). Further progress will rely mostly on improving TMR and enabling multi-level cell solutions for denser array designs [337], [343], [344].

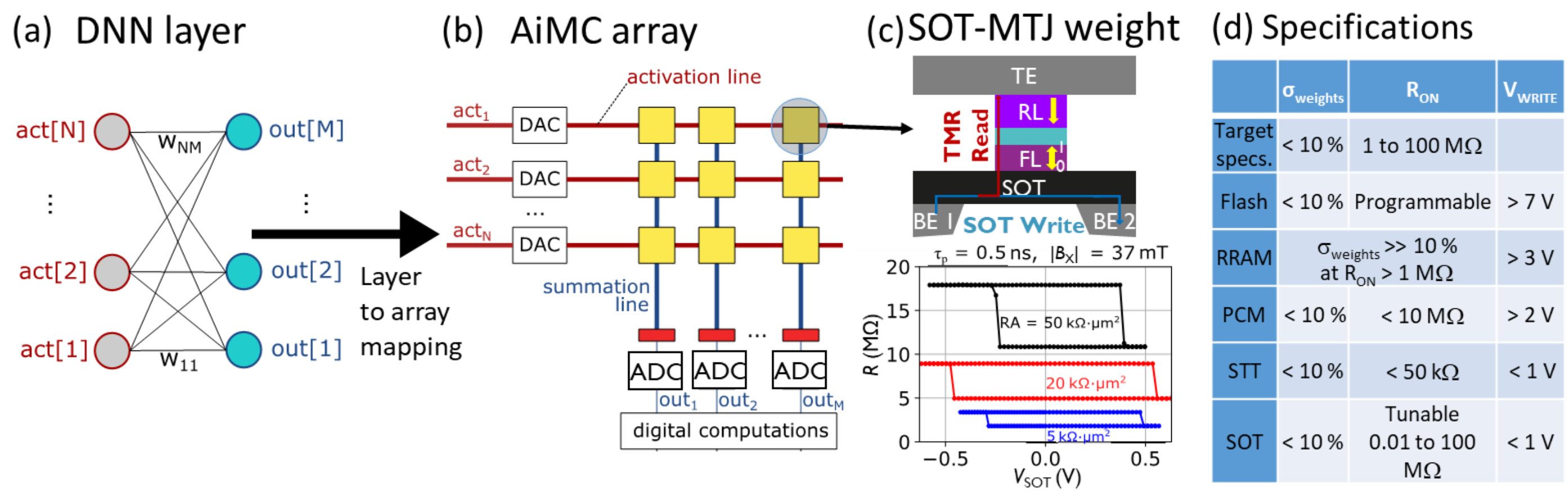

**Figure 13**. a) Deconvolutional Neural Network (DNN) layer implemented as b) an analog vector-matrix-multiply mapping array, c) SOT-MRAM cell used as highly resistive weight memory required for analog in-memory computing (AiMC), where DAC and ADC are digital-to-analog and analog-to-digital converters, respectively. d) Simplified comparison of different weight memories for analog DNN inference. Stated numbers represent typical operating conditions and are not fundamental lower limits. RON depends on AiMC size. RL: reference layer; FL: free layer; TE: top electrode; BE: bottom electrode; PCM: phase change memory.

As an extension of digital memory, SOT-induced magnetization control can be applied to realize artificial synapses or memristors and neurons [337]. The basic concept of memristor was originally introduced by Chua as the fourth circuit element following the resistor, inductor, and capacitor [345]. In 2008, Strukov *et al*. pointed out that their Pt/$TiO_{2-x}$/Pt device functions as a memristor [346]. In general, memristors represent devices whose conductance continuously changes with respect to the total amount of applied charge and which store the state in a long period, as shown in Fig. 14(a). The reason that the memristor has attracted great attention is its capability to be used in neuromorphic computing. While conventional computing hardware is an essential building block of today's information technology, it is recognized that there are several computational tasks such as cognition and inference that conventional computers cannot address efficiently. As a result, new computing hardware that is inspired by information processing in the human brain has been extensively explored. Fundamental units of the human brain are neurons and synapses that constitute a neural network. Memristors can be used as artificial synapses in artificial neural networks due to its similar functionality. Consequently, artificial synapses have been developed in various material systems, including resistance change systems, phase change systems, ferroelectric systems, and spintronic systems [347].

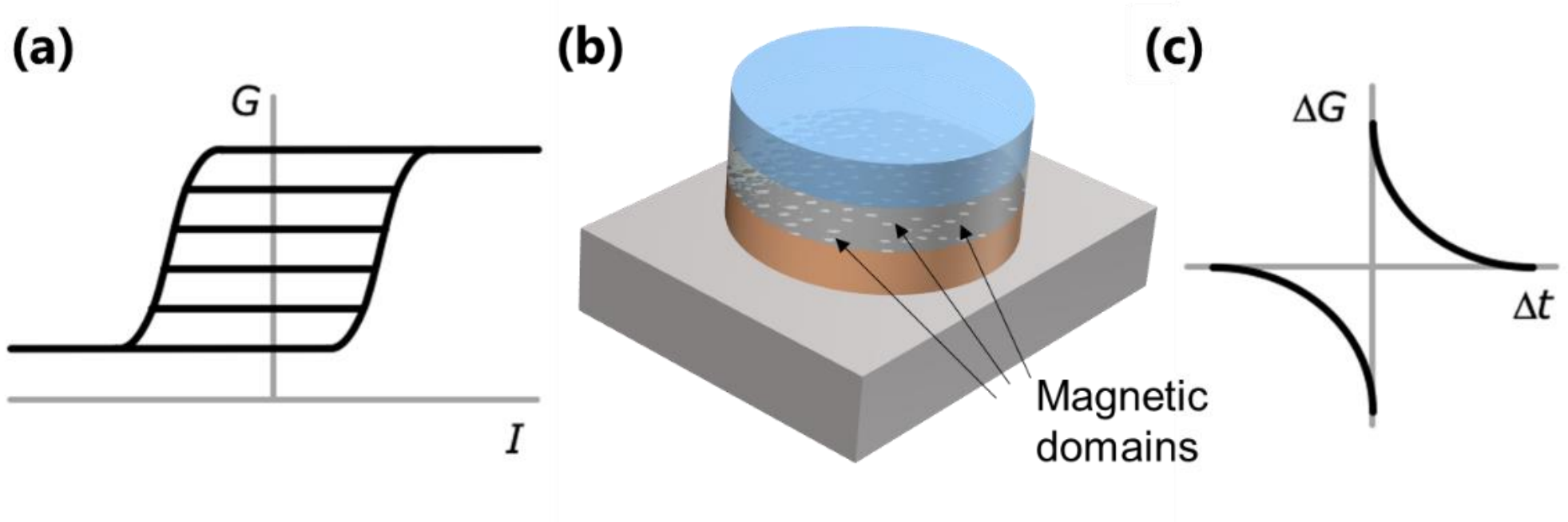

**Figure 14**. Functions of SOT memristor. (a) Conductance *G* versus current *I* of analog memory. (b) Snapshot of fine ferromagnetic domains in an antiferromagnet/ferromagnet system. (c) Conductance change D*G* versus timing difference between firing of post and pre neurons D*t* for spike-timing-dependent plasticity (STDP) synapse.

Memristive function in SOT switching was found in an antiferromagnet/ferromagnet system, where field-free switching is realized due to the exchange bias [207], [209]. A subsequent study revealed that the memristive behavior originates from the fact that magnetization switching proceeds with a unit of fine ferromagnetic domains which have different exchange biases resulting in different switching currents with respect to each other (Fig. 14b) [348], [349]. Subsequently, similar memristive behavior was observed in topological insulator/Mo/CoFeB [129], Ta/CoFeB/MgO [350], and Pt/ferromagnet/Ta/ferromagnet systems [351]. Also, current-induced manipulation of the Néel vector through the spin-orbit interaction in pure antiferromagnetic materials such as CuMnAs mostly shows memristive behavior [216], [218], [219], [352]–[354]. Furthermore, spike-timing-dependent plasticity (Fig. 14(c)), a dynamic property of biological synapse, was demonstrated in SOT memristors [351], [355], [356], showing promise for applications to asynchronous spiking neural networks.

Proof-of-concept demonstration of artificial neural networks using the SOT memristor as an artificial synapse was reported by Borders *et al*. [357] They used PtMn/[Co/Ni] systems and demonstrated an associative memory operation based on the Hopfield model [358]. In the Hopfield-model-based associative memory, several patterns are memorized as a synaptic weight of the artificial synapses and, once an input is given, the closest pattern is associated. Because the artificial synapse is required to store analog synaptic weight and is updated many times during the learning process, SOT memristor fits well with this application. Borders *et al.* confirmed a learning ability of the artificial synapse made of PtMn/[Co/Ni] and demonstrated fundamental operation of associative memory.

In order to make the SOT memristor a viable device in neuromorphic computers, one needs to substantially improve several device properties. One of the most important challenges is a reduction of device size while maintaining the memristive properties. As mentioned above, the memristive behavior of SOT devices is attributed to a separate switching of fine magnetic domains. Therefore, the reduction of device size in principle results in a decrease in the number of levels the device can show [348]. Meanwhile, in neuromorphic computing hardware, a large number of artificial synapses is required to be implemented, meaning that the memristor size is desired to be sufficiently small. To this end, the engineering of domain and domain wall structures is an important challenge. Also, to be used as artificial synapses, the resistance of MTJs should be sufficiently high, *e.g.*, higher than megaohms, because multiply-accumulate operation is performed in most of artificial neural networks where a large number of synapses are accessed simultaneously. In addition, most of the requirements for SOT-MRAM described earlier hold true for SOT memristor, including low critical current, large TMR ratio, high endurance, and compatibility with CMOS process. However, requirements for operation reliability are expected to be relaxed because artificial neural networks have redundancy to compensate for device unreliability.

Besides synapses, we also need the basic building blocks, *neurons*, for neuromorphic systems. A biological neuron resembles the leaky-integrate-fire dynamics, which accumulates the membrane potential as the input spikes arrive, with a leak factor. It has been shown that the magnetization dynamics in the SOT-MTJ, as well as the domain wall position in an SOT-based domain wall device, follow the leaky-integrate-fire behavior, thereby directly mapping the desired neuron characteristics to the device [355]. Interestingly, the same SOT-MTJ device can also be realized as a synapse, due to the inherent non-volatility of the magnetization. SOT-MTJs can be used as single-bit binary synapses, while SOT-based domain wall devices can store multiple bits per cell. The data can be written and read in the same fashion as in the memory application. However, connecting such SOT-MTJs in a crossbar-type fashion allows direct computation of the synaptic current, performing a highly-parallel matrix-vector multiplication operation by applying a voltage corresponding to the input at the rows and accumulating the currents along the column. Recently, neurons' spike-timing-dependent plasticity function has also been experimentally demonstrated (Fig. 14c) [355].

While SOT-induced magnetization switching has been extensively studied in the deterministic regime, the stochastic regime also holds enormous potential for application in various fields including neuromorphic systems, unconventional computing hardware, and information security. The usefulness of the stochastic nature of physical systems for computationally hard problems was pointed out by Richard Feynman in 1981 [359]. Recent studies have revealed considerable potential of spintronic systems for this purpose due to their well-controllable stochasticity. For STT-MTJ devices, true random number generation [360], [361], population coding [362], combinatorial optimization [363], and invertible logic [363], [364] have been demonstrated. For SOT devices, device-level investigations have also been initiated and various application areas have been proposed, including true random number generator [365], [366], Bayesian networks [367], [368], invertible logic [369], hardware security [370], and artificial synapses [371].

There are two types of approaches to utilize the stochasticity. The first approach (Fig. 15a) uses the probability of SOT-induced switching in binary devices, where the inherent time-varying thermal noise randomizes the magnetization dynamics [367], [368], [371]. In the stochastic regime, when a current is passed through the heavy metal, the magnet switches with some probability, which demonstrates a sigmoidal behavior as a function of the applied current. This stochastic switching can be leveraged to implement stochastic neurons and synapses, where analog information is encoded in probability, while the device itself is binary. The other approach (Fig. 15b) utilizes the superparamagnetic regime, where nanomagnetic devices designed to show fast thermal fluctuation between 0 and 1 states are employed [365], [372]. Here, SOT controls the ratio of time spent in the 0 and 1 states, allowing the devices to function as a binary stochastic neuron that is useful in stochastic neural networks and machine learning [373].

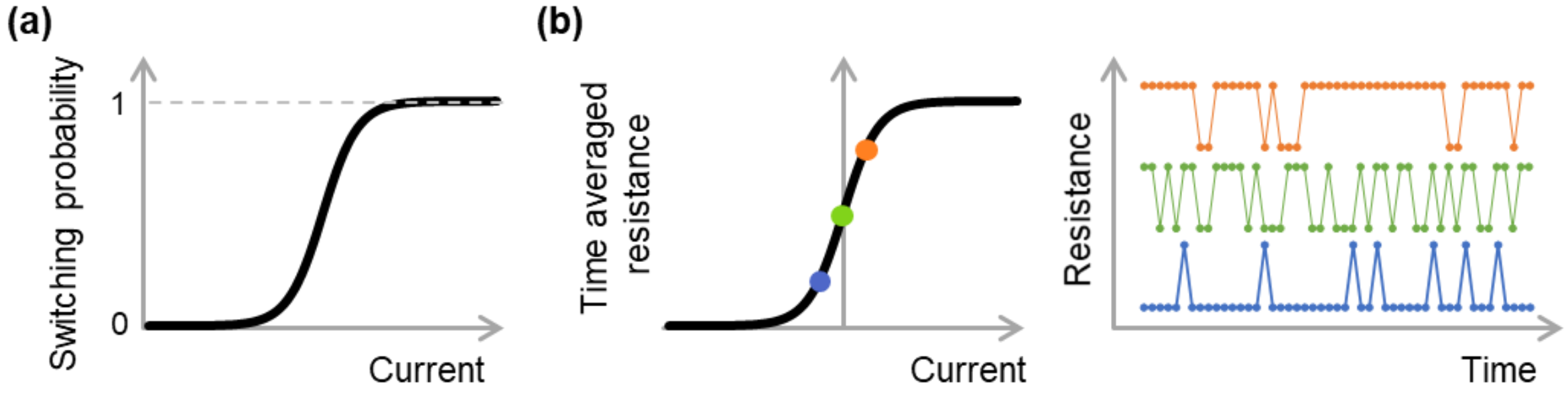

**Figure 15**. Two approaches to utilize stochastic behavior of SOT devices. (a) The first approach, where switching probability of binary devices is utilized. (b) The second approach, where the thermal fluctuation of superparamagnetic devices controlled by SOT is utilized. The right graph shows the temporal change of device resistance at three levels of applied current, indicated by closed circles with the same color in the left graph.

Several issues need to be addressed to effectively harness the stochastic behavior of SOT devices. First, physics to describe the probabilistic behavior under SOT should be accurately understood. While the probability of STT-induced magnetization switching has been well studied [266], [374], [375], few studies focusing on the probabilistic behavior of the SOT-induced switching have been performed. Second, for the approach using the superparamagnetic devices, acceleration of magnetization fluctuation is highly desirable to enhance the computational speed while maintaining reliability. For SOT devices, the fluctuation timescales reported so far range from milliseconds [372] to seconds [365], whereas the shortest timescale of STT-MTJs is 8 ns at negligible field/current bias [376] and 2 ns at high current bias [377]. In addition, robustness against external field disturbance and countermeasures against device-to-device variability may pose an important challenge as the stochastic/probabilistic behavior of SOT devices is expected to be sensitive to these factors. Third, the low $R_{on}/R_{off}$ ratio of these devices and process variations leading to reduced sense-margins is still a challenge for implementing such primitives at scale, and the research is ongoing.

## d. Field-free switching

Deterministic switching of perpendicular magnetization at zero external magnetic field is critical for SOT devices. In a typical SOT heterostructure, the spin-polarization is in the in-plane direction. Therefore, it is applied to both up and down magnetization states of the magnetic memory layer symmetrically. In order to realize deterministic SOT switching of the magnetic layer, this symmetry needs to be broken. In the earlier works, this symmetry was broken by applying an in-plane magnetic field collinear with the current. However, the presence of an external field is not practical for applications as it adds to the complexity of the device structure and operation. In recent years, there have been many efforts to realize deterministic field-free SOT switching of perpendicular magnetization using various approaches. Here, we have grouped these efforts into four categories of symmetry breaking: (i) **structural asymmetry**, (ii) **built-in in-plane fields**, (iii) **out-of-plane spin polarization**, and (iv) **hybrid approaches** (see Fig. 16).

**(i) Structural asymmetry**: This approach is based on introducing a lateral symmetry breaking in the device structure, typically achieved by using a wedge-shaped layer or by controlling the geometry of the magnetic layer. The wedge-shaped layer has been shown to induce either an effective out-of-plane field-like field [330], [378]–[383], or tilted anisotropy [384]–[386]. It has also been shown that out-of-plane field-like fields may even be present in the single-domain limit [383], which shows that this approach is potentially scalable. Furthermore, using structural asymmetry does not restrict the choice of material systems, nor does it add to the thermal stability requirements. However, the fabrication of a reproducible wedge layer is more of a serious challenge in practical large wafer-scale fabrication and thus should be mitigated for its practical use. In a system with gradient magnetic properties, the chiral symmetry of the SOT-generated spin textures can be broken by the DMI, and thus contributes to the deterministic field-free switching [387]. Alternatively, the controlled shape of the free layer relies on creating specific deterministic nucleation centers allowing for bipolar switching [388]. This approach is a credible field-less solution, readily integrable, but that would, however, face lithography and scaling limitations. A field-free SOT switching can also be realized by exploiting the domain wall motion in a dumbbell-shaped Ta/CoFeB/MgO nanowire with perpendicular anisotropy [389]. Due to DMI, the SOT switching was initiated by reverse domain nucleation at one edge followed by domain expansion across the nanowire. It was known that SOT induced domain wall motion does not require any external magnetic field in a nanowire with perpendicular magnetic anisotropy [390]–[393]. Such a simple Ta/CoFeB/MgO structure to realize the SOT-domain wall motion switching can be readily integrated with a conventional magnetic tunnel junction. The reliable and deterministic control of a single domain wall injection and domain wall displacement is obtained simply by sweeping currents, which enables a field-free SOT switching under repeated bipolar currents. The results are explained by the combination of the SOT-domain wall motion, the DMI, [394], [395] and the geometric domain wall pinning due to surface tension. [396] The proposed scheme can be easily integrated into three-terminal memory devices.

**(ii) Built-in in-plane fields**: This method is based on replacing the external in-plane field for deterministic switching with a built-in in-plane field/magnetic layer. This can be achieved by either an in-plane exchange bias field (through an antiferromagnet interface) [207]–[210], [397], an additional in-plane magnetic layer [331], [398]–[401], or chiral coupling to an in-plane magnet [402]. One problem with the in-plane exchange bias approach is that the exchange bias field has a non-uniform distribution, which results in incomplete magnetization switching [209]; reducing the device dimensions to single-domain scales may alleviate this issue [348]. Furthermore, it has been shown that exchange bias can be susceptible to current-induced Joule heating during the SOT switching process [397]. Another potential problem with this approach is that it limits the material choice to only a few antiferromagnets, which do not necessarily have high SOT efficiencies and often require larger thicknesses of the SOT channel, penalizing the write current. Using a built-in in-plane magnetic layer is a compact and straightforward solution. Still, it would impose a significantly larger write current as part of the current is shunted in the underneath in-plane magnetic layer. In addition, the torque applies to both magnetic layers causing a possible reversal of the in-plane ferromagnet and therefore high write error rates. Finally, having a separate in-plane magnet does not have these problems and is among the best potential candidates for practical applications as recently demonstrated [324], [403], [404]. The same performances as standard devices are kept, the in-plane field amplitude can be tuned independently of SOT-MTJ stack, and the scalability is similar to three transistor (3T) standard cells. However, having an additional magnet will require precise control of its properties, orientation, and retention to avoid unintentional writing error rates.

**(iii) Out-of-plane spin-polarization**: Out-of-plane spin-polarization can break the symmetry between the up/down magnetization states, resulting in field-free SOT switching. This has been achieved (1) in crystalline $WTe_2$ [61], $IrMn_3$ [203], $Mu_2Au$ [405], and CuPt/CoPt interface [406], using lack of lateral inversion symmetry in its crystal structure, where an out-of-plane spin polarization can be generated by the in-plane current; (2) in an in-plane CoFeB/Ti/perpendicular CoFeB system, where the spin precession by the interfacial Rashba field induces the out-of-plane component of the spin polarization [43]; and (3) in a heavy-metal bilayer with the opposite SOT efficiencies or spin Hall angles, where the competing spin currents generate an out-of-plane spin polarization [407]. Among these approaches, the heavy-metal bilayer system is potentially a practical solution for device applications with high energy-efficiency, provided the confirmation and optimization of this mechanism.

**(iv) Hybrid approaches**: Spin-orbit torques can also be combined with other voltage- or current-induced effects for realizing field-free switching. Recently, field-free SOT switching has been realized in a ferroelectric/ferromagnetic structure using PMN-PT [23], and by combining spin-transfer torques and SOTs [329]. The challenges for these approaches are that the addition of a ferroelectric layer adds to the stack complexity, and using spin-transfer torques potentially limits the switching speed and device endurance.

We compare different aspects for all these schemes of field-free SOT switching in Table 1.

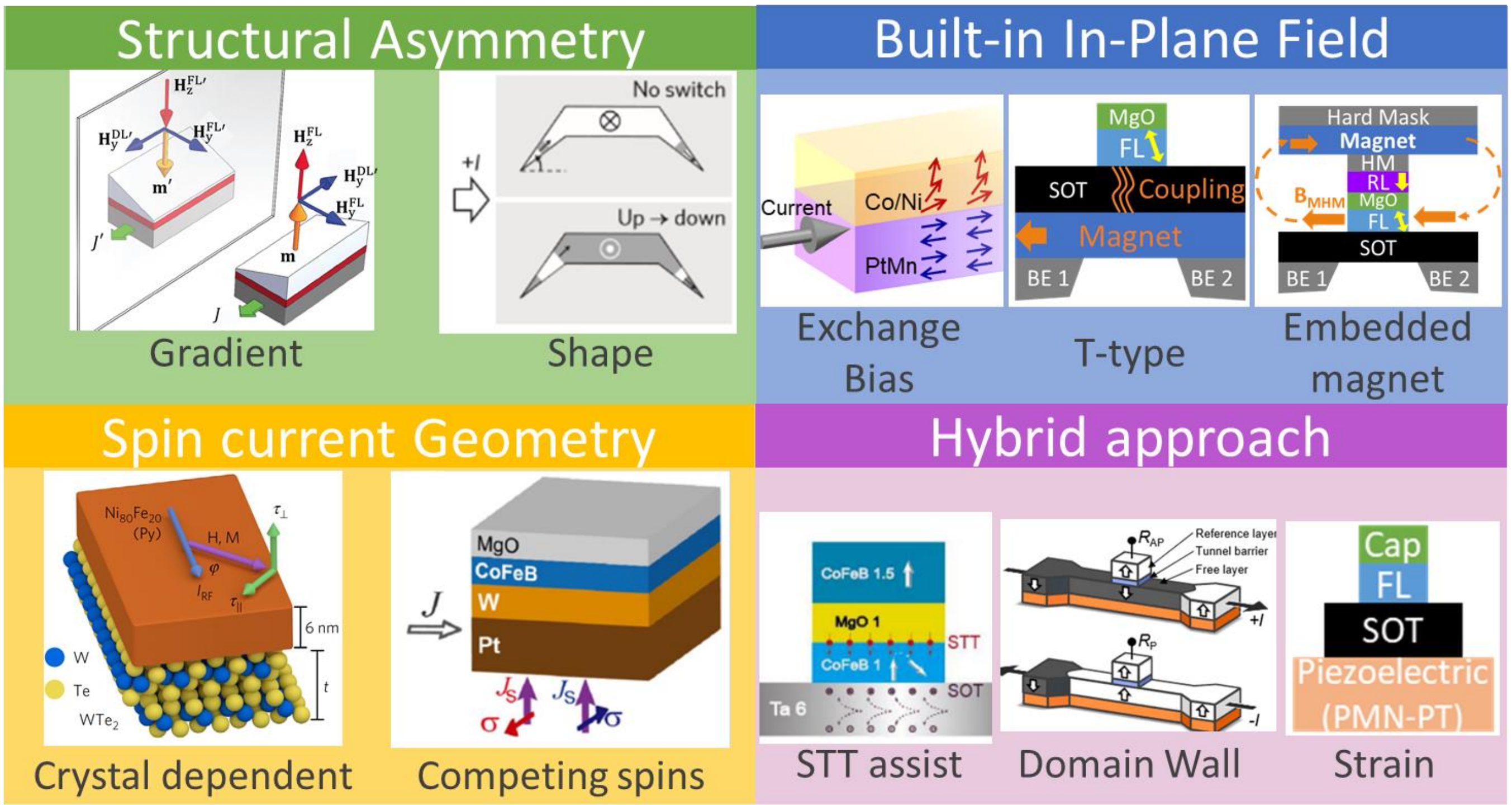

**Figure 16**. Different approaches for field-free SOT switching. (i) Structural asymmetry, (ii) built-in in-plane fields, (iii) out-of-plane spin polarization, (iv) hybrid approaches. Figures are adapted from refs. [39], [43], [23], [207], [329], [330], [388], [389], [407]. "Shape" figure is reprinted with permission from [388]. Copyright (2016) by the Springer Nature. "Crystal dependent" figure is reprinted with permission from [39]. Copyright (2017) by the Springer Nature. "Competing spins" figure is reprinted with permission from [407]. Copyright (2018) by the American Physical Society. "Domain wall" figure is reprinted with permission from [389]. Copyright (2018) American Chemical Society.

**Table 1.** Comparison between various schemes to achieve field-free SOT switching

| | **Built-in in-plane field** | | | **Structural asymmetry** | | **Hybrid approach** | | | **Spin Current Geometry** | | |
|---|---|---|---|---|---|---|---|---|---|---|---|
| | Embedded magnet | T-type or in-plane polarizer | AFM with EB | Shape | Gradient | STT assist (PMA free layer) | Domain wall | Strain | In-plane MTJ | Competing spin current | Crystal symmetry (include 2DM) |
| **Maturity** | IV: MTJ wafer scale | II: Micromagnet | II: Nanomagnet | II: Micromagnet | I: Hall bar | IV: MTJ wafer scale | I: Wire | II: Micromagnet | IV: MTJ wafer scale | I: Hall bar | II: Micromagnet |
| **Reported minimum free layer size** | 60 nm (circular) [324] | 4 µm x 4 µm [43] | 50 nm (circular) [348] | ≈ 3 µm x 1 µm (U or S shape) [388] | 300 nm (channel width) [383] | 57 nm (circular) [408] | 5 µm (channel width) [389] | 3 µm (circular) [23] | 88 nm x 315 nm (XY-type) [325] | 20 µm (channel width) [407] | 5 µm (circular) [406] |
| **Reported minimum write pulse duration** | 0.3 ns [324] | 20 µs [43] | 1 ns [355] | 35 ns [388] | 10 ns [383] | 10 ns [408] | 1 ms [389] | 1 s [23] | 0.2 ns [409] | 50 µs [407] | 30 µs [406] |
| **Limitations** | Additional magnet | Current shunt; Coupling mechanism; Dynamic interplay IP/OOP layers | EB distribution; Thickness, EB vs. $T_b$ | Scalability; patterning | Feasibility at nanoscale? | Endurance/ reliability; Dominant mechanism; No R/W decoupling | Controllability ; Scalability | Integration scheme; Cell size | Cell Size (ellipse) | $I_{sw}$ / Efficiency?; Further confirmations | Thickness, crystallinity control (substrate); 2DM: growth, transfer |

Note: Maturity of research stage is defined as four levels, I: Hall bar or Wire (switching device), II: Nanomagnet or micromagnet (pattern on channel), III: MTJ, and IV: MTJ wafer scale. Examples of these schemes: embedded magnet

[324], in-plane-polarizer [43] or T-type [401], antiferromagnet (AFM) with EB: exchange bias [207], [348], [355], shape [388], gradient [330], [383], STT assist [329], [408], domain wall [389], strain [23], in-plane MTJ [7], [325], [409], competing spin current [407], crystal symmetry (include 2DM: 2D materials) [39], [406].

## e. Terahertz generation using SOT

Since ultrafast demagnetization at the time scale of sub-picosecond was first observed using a femtosecond laser pulse in 1996 [410], the field of terahertz spintronics has raised a lot of interest not only in characterizing magnetic materials [411], [412], but also in generating terahertz with spintronic devices [413], [414]. The lack of efficient, low-cost, and broadband terahertz sources is one of the bottlenecks of wide spreading terahertz technologies, especially for terahertz time-domain spectroscopy.

In the first demonstration of terahertz emission from the Fe/Au bilayer structure (Fig. 17), a weak terahertz amplitude, 100 times smaller than a standard ZnTe terahertz emitter, was observed [414]. Such an FM/nonmagnetic (NM) heterostructure needs to be excited by a femtosecond laser pulse. The laser pulse excites an ultrafast spin current in the ferromagnetic layer, which then diffuses to the nonmagnetic layer. Transient charge current is then generated in the nonmagnetic layer based on the inverse spin Hall effect or inverse Rashba effect, followed by the radiation of terahertz waves.

Since the first demonstration, ultrafast spin-to-charge conversion and the associated terahertz emission has been explored in various magnetic heterostructures [415]–[420]. Furthermore, a recent unveiling of the ultrafast spin-to-charge conversion at the time scale of 0.12 ps in magnetic heterostructures opens up a route to potential spintronic devices manipulating spin currents on terahertz timescale [421]. Recent advances show that a low-cost magnetic thin film deposited by sputtering can generate broadband terahertz radiation with an intensity comparable or even stronger than the standard ZnTe terahertz emitters with a conversion efficiency > $10^{-4}$ [415]–[419]. Moreover, this efficient terahertz source can be noise resistive, magnetic field controllable, flexible, robust, and low-power fiber laser-driven. [417] These results pave the route for developing efficient terahertz sources based on magnetic heterostructures.

Terahertz emission from more complex magnetic heterostructures including nearly compensated ferrimagnetic alloys or antiferromagnet layers were studied. In previous studies utilizing FM/NM bilayer structures, the spin current generation is attributed to the net magnetization in the ferromagnetic layer [414], [417]. However, it is found that the emitted Terahertz field is determined by the net spin polarization of the laser-induced spin currents rather than the net magnetization in ferrimagnetic layers. [422], [423] On the other hand, an antiferromagnetic layer plays a different role in terahertz emission. For example, antiferromagnetic material (e.g., IrMn) is not a good spin current generator by a femtosecond laser excitation, but it is a good detector. [422] These results not only suggest that a compensated magnet can be utilized for robust terahertz emission but also provide a new approach to study the magnetization dynamics, especially near the magnetization compensation point.

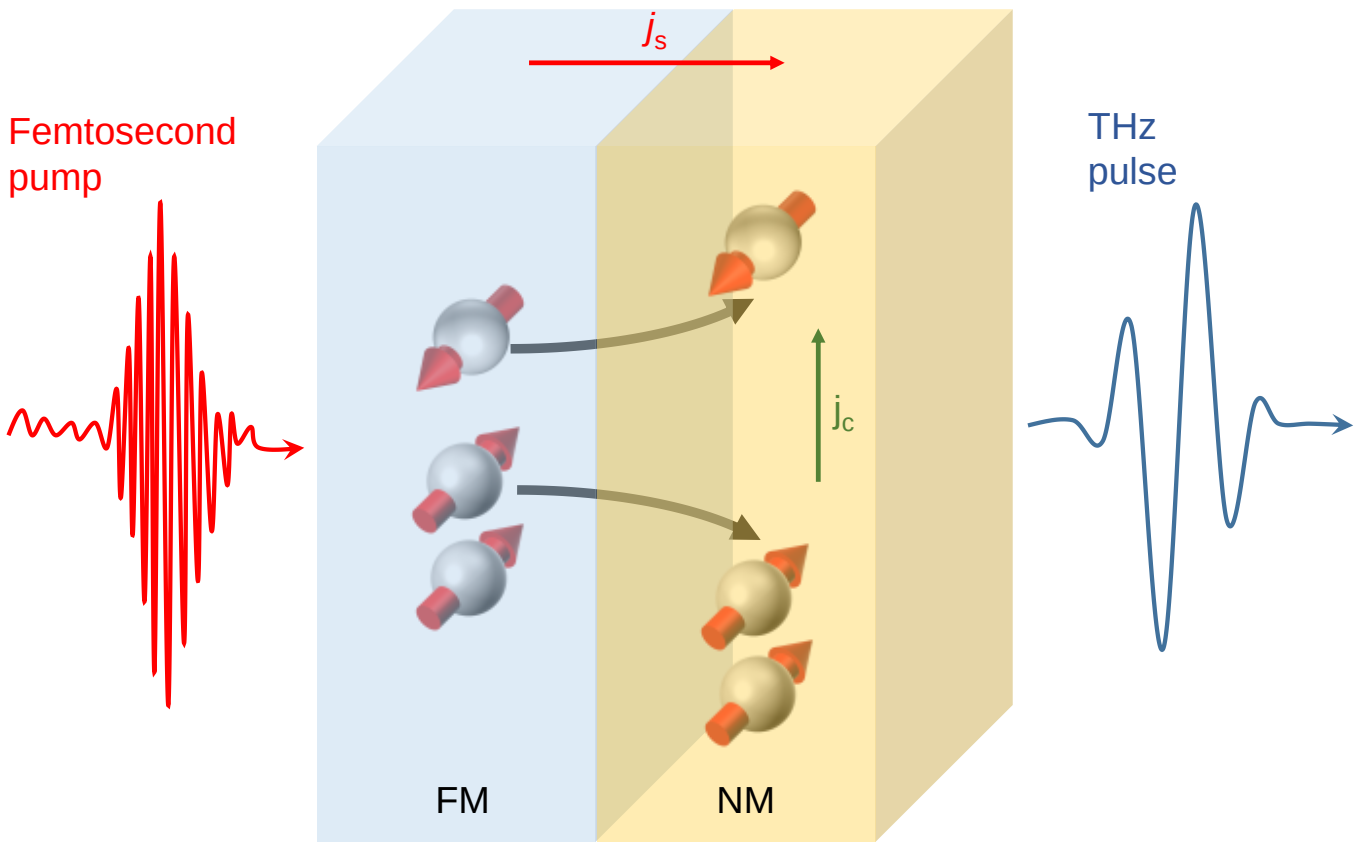

**Figure 17**. Spintronic terahertz emitter.

The nanometer thicknesses found in spintronic devices allows for easy phase matching with other terahertz emitters, and a hybrid terahertz emitter combined with semiconductor materials can be realized [424]. With a bias current, there is a contribution of terahertz emission from semiconductor materials, which can be constructively interfered with the terahertz signals generated from the magnetic heterostructures. Consequently, a two to three order enhancement of the terahertz signals was achieved in a lower terahertz frequency range (0.1 THz to 0.5 THz), in which ferromagnetic/nonmagnetic heterostructures show relatively poor performance. In addition, the performance of this hybrid emitter at higher frequencies is comparable to the FM/NM heterostructures. These findings push forward the utilization of spintronics based terahertz generation devices for ultra-broadband terahertz applications.

In the future, terahertz time-domain spectroscopy can play an essential role in the characterizations of novel materials, especially ultrafast spin dynamics [425]. Recently, Weyl semimetal materials (e.g., TaAs and $WTe_2$) have emerged due to the intrinsic property of strong spin-orbital coupling, which is significant for the generation and detection of spin currents. Such exotic large spin-orbit coupled systems are interesting materials for ultrafast spintronic devices as well as terahertz generation devices [426].

## f. SOT nano-oscillators

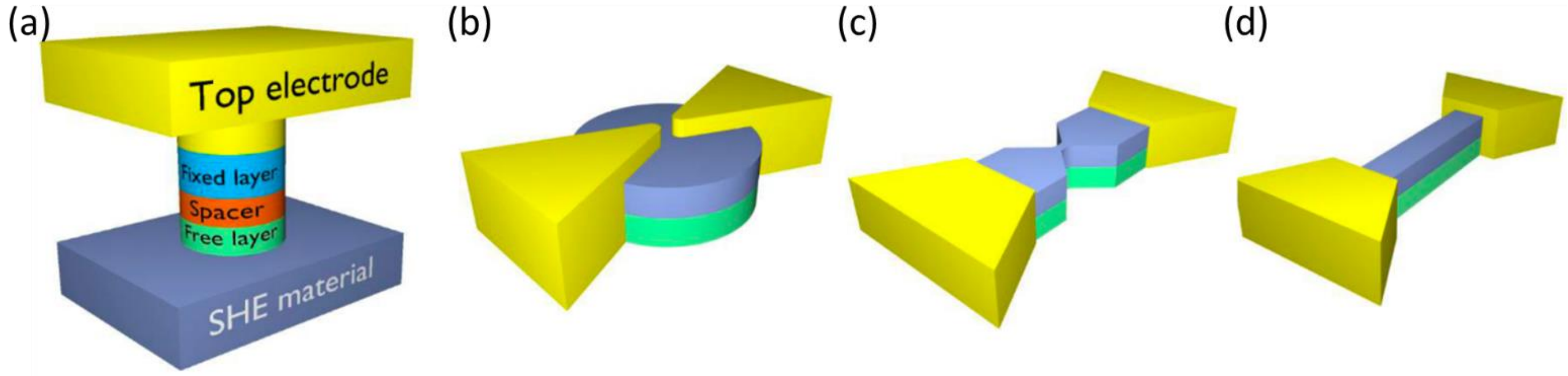

**Figure 18**. Device schematics of SOT nano-oscillators. (a) nano-pillar; (b) nano-gap; (c) nano-constriction; (d) nanowire

Spin Hall nano-oscillators (SHNOs) are microwave signal generation devices where a pure spin current from a material with SOT drives sustained auto-oscillations of the local magnetization of a nearby ferromagnetic layer [277]. SHNOs have been fabricated using a wide range of different layouts (Fig. 18) such as ferromagnetic nano-pillars on extended heavy metal films [427], extended HM/FM bilayers with Au nano-gap electrodes [428], [429], HM/FM nanowires – either uniform [430], tapered [431], or with thickness variations [432] – and extended HM/FM films with lateral nano-constrictions [433]. Given sufficient intrinsic SOT of the ferromagnet or its interfaces to adjacent insulators, SHNOs have also been demonstrated without a heavy metal layer. The presence and dynamics of a SOT driven domain wall can also be used as an SHNO [434]. With the exception of nano-pillars, all SHNOs have easy optical access to the magnetodynamically active regions, which allows for direct microscopy using magneto-optical Kerr effect [435]–[437] and Brillouin Light Scattering microscopy [428], [437]. Nano-constriction SHNOs have been fabricated down to 20 nm [438], operated up to 26 GHz [439], and can be made on Si substrates and materials entirely compatible with CMOS [439]. They can also be frequency modulated via their drive current. [440] Depending on their layout, the choice of material properties, and the applied field direction and magnitude, a large number of auto-oscillating spin-wave modes have been observed. These include spin-wave bullets [428], magnetic droplets [441], edge modes [433], and propagating spin waves [432], [442]. The latter are realized using perpendicular magnetic anisotropy, and since this property can be voltage-controlled, gated SHNOs have been demonstrated with voltage-controlled auto-oscillation frequency [443]. Since the spin waves in such devices can be continuously tuned from a localized to a propagating mode, where the latter experiences dramatically increased dissipation from the radiated spin waves, even a modest frequency tunability can lead to a giant change of the effective SHNO damping and the threshold current, such that the voltage can be used to turn the SHNO on/off at constant drive current [443].

Nano-constriction SHNOs have very recently also shown a great propensity for mutual synchronization, both in one-dimensional chains [444], [445] and in two-dimensional arrays [446]. As predicted by theory [447], [448] and shown experimentally, the coherence of the microwave signal grows linearly with the number of mutually synchronized constrictions. As the phase noise is one of the main limiting factors for SHNO applications, the close to two orders of magnitude improvement in mutually synchronized arrays of 64 SHNOs shows great potential for the future. Mutual synchronization can also be used for different types of neuromorphic oscillator computing [449], including Ising machines [450], and two-dimensional SHNO networks appear to be ideally suited for such applications [446]. Using individually voltage-controlled SHNOs in large arrays, with the added functionality of integrated memristors, is one particularly promising route [451].

Future research into SHNOs and their networks should be carried out along a number of different dimensions such as power consumption, frequency range, output power, network size, network topologies [452], and improved local tunability of individual spin Hall nano-oscillators in large networks.

## g. SOTs with domain walls and skyrmions

*SOT driven chiral domain wall motions*: Magnetic domains are small magnets that are uniformly magnetized inside. When neighboring domains are magnetized in different orientations, there are boundaries between domains that are named magnetic domain walls.  Besides magnetic skyrmions [453], [454], domain walls are topological objects such that they are topologically protected, thereby allowing domains to encode data in memory and logic. To access the data that are encoded by domains, the domain walls must be moved, which can be achieved by electrical current pulses, thus showing a sharp contrast with field-driven domain wall motion. Electrical current pulses move the domain walls along the electron flow direction by spin-transfer torque (STT), irrespective of domain wall configurations, owing to angular momentum conservation since the spin current is repolarized in each domain. On the other hand, for the field-driven case, the domain walls move in opposite directions depending on the domain wall configurations to minimize the Zeeman energy, thereby annihilating all data. Although STT driven domain wall motion provides the fundamental mechanism for domain wall-based devices, the access time is limited by an intrinsic domain wall pinning [455], [456], Walker-breakdown [457], and low domain wall mobility [458].

Meanwhile, SOT [7], [8] has emerged as a possible route to resolve the limitation in STT-driven domain wall motions. In this scenario, the chiral domain walls are stabilized by anti-symmetric exchange interaction - DMI [391], [392]. DMI is typically found in systems with broken inversion symmetry, like interfaces. In addition, the interface-induced DMI favors Néel-type walls in which the domain wall magnetization is out of the wall plane. To stabilize Néel-type walls against Bloch-type walls that favor domain wall magnetization in the wall plane, the interface-induced DMI induces localized DMI fields oriented in opposite directions for the different domain wall configurations. The DMI fields act to move the chiral domain walls as the SOT assists in rotating the magnetization away from the Néel wall direction, thereby giving rise to domain wall tilting. The SOT and interface-induced DMI determine the sign and strength of domain wall velocity significantly higher than STT exhibiting no Walker breakdown or intrinsic pinning. The chiral nature leads to the same velocity (sign and magnitude) for both domain wall configurations, while the signs of domain wall tilting angles are opposite.  Note that the domain wall tilting decreases the efficiency of SOT driven chiral domain wall motion. Hence, if the chiral domain walls are injected into inversion symmetry broken wires like Y-shaped [459] or curved ones [460], the domain wall displacement or velocity becomes asymmetric between the branches in the Y-shaped wire. The signs of curvature in the curved wire depend on domain wall configurations. This may potentially limit the increase in density of chiral domain wall-based memory.

The asymmetry of domain wall velocities between the domain wall configurations can also be induced by magnetic fields along the wire direction that are collinear with the DMI fields [392], [461]. The applied field compensates or adds to the DMI fields depending on the domain wall configurations with a given applied field. Consequently, the velocity vs. field curve is mirror-symmetric with respect to zero fields for two domain wall configurations. This shows that the current-driven chiral domain wall motion is susceptible to possible external fields like the stray field or Oersted field, thus limiting the development of robust chiral domain wall based devices. The solution to this challenge can be found from antiferromagnetically coupled composite chiral domain walls as discussed below.

*Devices based on SOT driven chiral domain walls*: The SOT driven chiral domain wall motions have many advantages over the STT counterpart such as high efficiency, tunability, and additional functionalities for applications. The chiral domain wall based devices that may be of interest are racetrack memory, neuromorphic devices, and logic. The racetrack memory [2], [462], [463] breaks down into multi-domain wall and single-domain wall (1-bit) three-terminal racetrack memories [464]–[466]. The multi-domain wall racetrack memory was proposed to replace hard-disk drives that require mechanical motion, thereby achieving a storage class memory with permanent endurance, large capacity and high speed. A proof of principle for the multi-domain wall racetrack memory was previously demonstrated with shift register operation by current pulses [467], and later the multi-chiral domain walls were shown to be moved by SOT [391], [392], [468], [469]. However, the 3D integration of chiral domain walls for further increase of the density faces some technical challenges, thus forming a long-term technology. On the other hand, the 1-bit three-terminal racetrack memory has risen to be a promising near-term non-volatile fast memory since it needs significantly less technical requirements than the multi-domain wall racetrack [389]. For example, the nucleation of a single domain wall in each device is required only once in a manufacturing step. Moreover, there is no worry about narrowing and securing wall-to-wall distances. The write speed in 1-bit three-terminal racetrack is determined by the domain wall velocity, while the on-off ratio for read-out typically relies on the tunneling magnetoresistance in an MTJ in which the domain wall track channel is used as a free layer for the data bit. Hence both the SOT and DMI need to be increased to enhance the chiral domain wall velocity. The 1-bit three-terminal racetrack can be used for neuromorphic functionality [462].

One of the biggest challenges in 1-bit three-terminal racetrack is to reduce threshold current density $J_C$ above which a chiral domain wall starts depinning. Typical $J_C$ values e.g., in Pt/Co are in the order of $10^7\ \mathrm{A/cm^2}$ [8], [458], which is required to be lowered by at least one order of magnitude for practical applications. The reduction of $J_C$ is partly correlated with anisotropy and SOT and DMI strengths since the larger domain wall width and the stronger SOT and DMI make it easier for domain walls to be depinned. Material exploration to achieve this have been extensively carried out. For example, recently metastable Pt assisted by Bi surfactant material has shown dramatic decrease in $J_C$ and an increase in the domain wall mobility with a given current density[470]. Thermal stability is another issue that can be a trade-off with the reduction of $J_C$ like STT-MTJ. Engineering pinning potentials of domain walls is required to lower $J_C$ while maintaining high thermal stability, like geometrical design of extrusion/constriction of racetrack wires. Synthetic antiferromagnets or ferrimagnets may be a good approach to resolve this challenge as discussed below.

The chiral nature of those domain walls can be utilized for logic gate applications. Recently all-electrical logic operations such as NAND, NOR, XOR and full ADDER have been replicated using chiral domain walls, which are based on the inversion of the domain walls, i.e. NOT-gate [402]. A key component in logic gates is the artificial chiral domain wall that is static and lithographically fabricated. As a normal chiral domain wall is injected into the static domain wall region, the outgoing domain wall is inverted by the interplay of domain wall chirality with magnetostatics [471]. When NOT-gates are combined to form a junction together with a bias, the output can be determined from the inputs according to the majority gate rule, thereby leading to NAND or NOR gate depending on the bias. Note that additional access to the bias allows the development of reconfigurable logic. However, there are technical challenges to commercialize the chiral domain wall logic chips, including issues of synchronization and feedback logic circuit.

*Current induced domain wall dynamics of antiferromagnetically coupled composite chiral domain walls*: We have discussed the limitation of the commercialization of chiral domain walls-based devices above. Meanwhile, it has been discovered that composite domain walls can be moved much faster than single-layer and ferromagnetically coupled ones. The two chiral domain walls are antiferromagnetically coupled across a spacer layer such as Ru via the Ruderman–Kittel–Kasuya–Yosida interaction, thus showing that the domain wall velocity is highly correlated with the exchange coupling sign [253]. The main driving force to move domain walls turned out to be a powerful torque, exchange coupling torque that is effective only in the antiferromagnetically coupled case. Moreover, exchange coupling torque increases as the net magnetization becomes compensated, thereby forming a maximum velocity at zero net magnetization. Similarly, enhancement of SOT was also observed in ferrimagnets near the compensation [245].

In contrast with the single-layer domain wall case, the velocity versus longitudinal field curves form a broad maximum around zero-field in the fully compensated antiferromagnetic composite domain walls, showing symmetry with respect to zero-field [253], [461], [472]. In addition, the curves for two domain wall configurations are nearly identical exhibiting insensitivity of domain wall dynamics to domain wall configurations in the presence of external fields. Importantly, the domain wall tilting in antiferromagnetic composite domain walls is found to be small. This is concluded from the observation that the asymmetries in antiferromagnetic composite domain wall displacement disappear when injected into Y-shaped and curved wires. This is owing to the compensation of the domain wall tilting in each component domain wall. They allow the antiferromagnetic composite domain walls to be even more useful for commercializing domain wall-based devices, as discussed below.

The antiferromagnetic composite domain walls can be extended to ferrimagnetic chiral domain walls that are typically formed from rare-earth and 3d transition-metal alloys or multilayers in which two networks (rare-earth and 3d transition-metal) are oriented in an anti-parallel way [472]. The net magnetization in such ferrimagnets is sensitive to the composition ratio and temperature since the magnetic moment per atom and Curie temperature for each network is different [239]. The difference in gyromagnetic ratios between rare-earth and 3d transition-metals gives rise to the disparity between angular momentum and moment compensations [245]. Interestingly, the current-driven domain wall dynamics turn out to rely on the angular momentum rather than the magnetic moment. Consequently, the domain wall velocity becomes maximized at the angular momentum compensation at which the exchange coupling torque forms a maximum, while the other torques vanish. Since the exchange coupling in ferrimagnets is much larger than the synthetic antiferromagnets, the ferrimagnet domain wall velocity by the current can be higher than 1 km/s [185], [254].

When the exchange coupling strength in an AF composite domain wall is weak, while the positions of component domain walls are tightly locked, an exotic domain wall dynamic can emerge [473]. Since the torques on component chiral domain walls are not identical, the faster domain wall drags the slower one, which is namely chiral exchange drag. As the disparity of the velocities is larger than a threshold value, the energy to move the composite domain wall is converted into the one that precesses/oscillates domain wall magnetizations. Thus, they lead to a dramatic reduction of composite domain wall velocity, which corresponds to chiral exchange drag anomaly. Note that the precessions/oscillations of chiral domain walls are all synchronized, and the frequency increases with the increasing disparity of velocities.

*Devices based on antiferromagnetically coupled composite and ferrimagnetic chiral domain walls*: antiferromagnetically composite domain walls have many benefits compared with single-layer domain walls for applications to racetrack memory and logic. First, more robust racetrack memory and logic are allowed, since the antiferromagnetic composite domain walls are significantly inert to possible external fields since the stray field is minimized. Second, the access and operation speeds are much higher due to the giant exchange coupling torque as the net magnetization is fully compensated. Third, the domain wall tilting and the asymmetries in domain wall velocity between two domain wall configurations by geometrical broken symmetries or external fields hinder the increase of density of racetrack memory as discussed above. Fourth, the on-off ratio for read-out in an MTJ can be large, since the magnetic layers that sandwich a tunnel barrier can have a large spin polarization. Fifth, the $J_C$ can be significantly reduced in antiferromagnetically coupled composite domain walls, due to the absence of domain wall tilting, small net magnetization, and additional room to reduce anisotropy, while maintaining decent thermal stability.

The chiral exchange drag anomaly from the weakly-coupled antiferromagnetic composite chiral domain walls can be used to develop a gigahertz range of oscillators that is tunable by external fields [473].

We present a roadmap for domain wall-based SOT devices in Fig. 19.

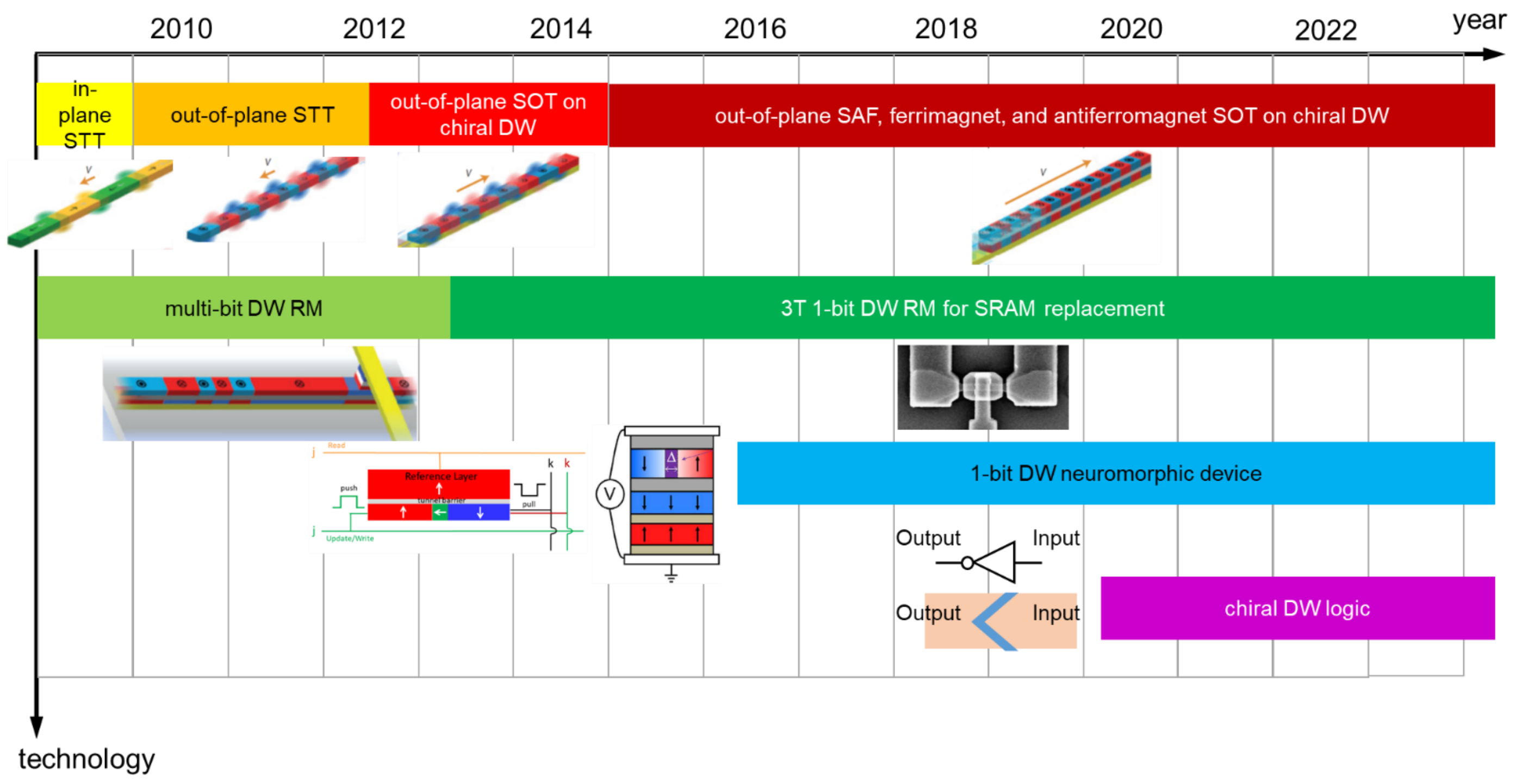

**Figure 19**. Roadmap of domain wall (DW)-based device technologies. Top bars show the evolution of domain and domain wall types and torques. The second, third and fourth bars from the top describe the racetrack memory (RM), neuromorphic devices and logic, respectively. The inserted figures are adapted from refs. [463], [474], [475].

*Skyrmions*: Another potential technology where SOTs play a crucial role is information storage and processing with magnetic skyrmions (Fig. 20). Magnetic skyrmions are topologically stabilized spin textures that can be translated effectively using SOTs [476]–[481]. The prospect of using these magnetic structures in efficient racetrack memories or for brain-inspired computing has sparked tremendous interest. Recently, it has been experimentally demonstrated that individual skyrmions can form at room temperature due to enhanced thermal stability in carefully designed magnetic superlattices, for instance, of Pt/CoFeB/MgO [478], Ir/Co/Pt [482] and Co/Pd [483] multilayers, where the stabilization of topological spin textures originates from the DMI.

Due to their non-trivial topology, skyrmions move under an angle with respect to the direction of the applied current [477], [479], [480]. The deflections from this skyrmion Hall effect complicate the use of skyrmionic spin structures in racetrack devices, e.g., for logic functionalities. Shaping the complex trajectory of topological spin textures relies on our microscopic understanding of the interplay between spin topology, damping, and current-induced SOTs [484]. While the Thiele equation of motion [485], [486] accounts for this interplay, the corresponding treatment has led to the common perception that the field-like SOTs are irrelevant for describing the dynamical properties of rigid skyrmions as well as the skyrmion Hall effect. Recently, experiments and micromagnetic simulations have suggested that damping-like SOTs can induce skyrmion deformation and thus cause a large skyrmion Hall effect [487].

The non-trivial form of SOTs in magnetic multilayers requires an extension of this widely accepted picture. While the current-induced dynamics of skyrmions have been explained in terms of deformations of the spin texture [479], the anisotropy of spin-orbit torques plays an essential role in predicting and interpreting the dynamics of topological spin structures [488]. Recent results show that the coupling of magnetic textures to field-like torques that are higher-order in the local magnetization can give large corrections to the skyrmion Hall effect. First-principles calculations and symmetry arguments quantify the relevance of these modifications for the dynamical properties of skyrmions and antiskyrmions in layered magnetic films of Ir/Co/Pt and Au/Co/Pt. They suggest that engineering the anisotropy of the SOTs provides a new perspective for controlling the skyrmion Hall effect and the motion of skyrmions or antiskyrmions in multilayer systems.

Recently DMI and electronic signatures of skyrmions in magnetic insulator-based heterostructures have been shown [184], [185], [489]. Domain walls and skyrmions in these magnetic insulators could promise even lower power dissipation of spintronic devices thanks to their low damping.

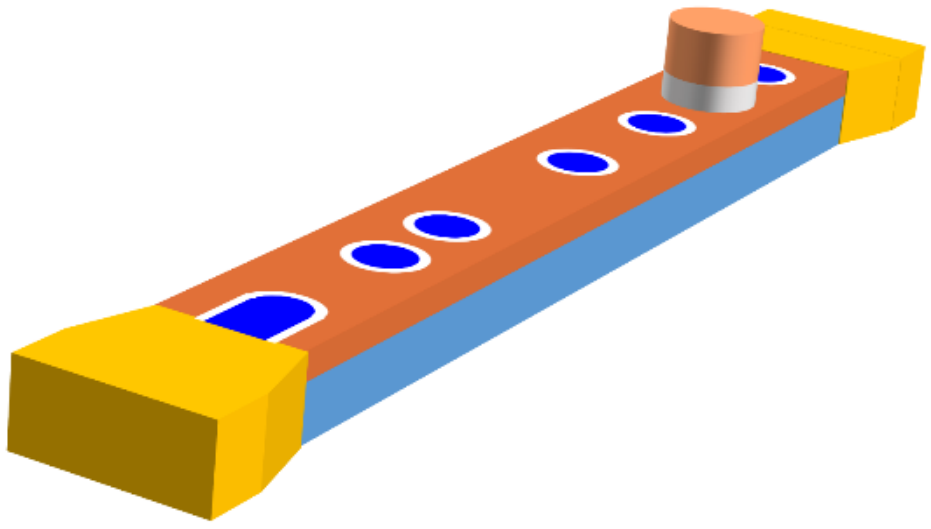

**Figure 20**. Skyrmion racetrack memory. The blue dots represent skyrmions and the information readout is done through the MTJ on top of the skyrmion racetrack.

## h. Industrialization considerations

Due to their superior scaling properties and non-volatility, STT-MRAMs are presently in the path of commercialization as a replacement for slow SRAMs and eFlash in embedded cache memories, with potential applications also as a persistent DRAM technology. Several integrated circuit companies as well as toolmakers are aggressively developing STT-MRAM technologies beyond the 22 nm node with chip capacities larger than 1 Gb. However, STT-MRAMs will be ultimately limited to the last level cache (L3/L4) due to the relatively large switching latency of STT and high currents required to reach nanosecond and sub-nanosecond switching times, which can damage the MTJ tunnel barrier. SOT-based MRAM allows for switching the free layer without passing a current through the tunnel barrier, thus minimizing the risk of voltage breakdown and offering unmatched switching speed and endurance compared to STT-MRAM. State-of-the-art experiments demonstrate reliable sub-nanosecond switching (300 ps) of three-terminal MTJs based on [Ta, Pt, W]/CoFeB/MgO with either perpendicular [324] or in-plane [325], [490], [491] magnetization in the complete absence of external fields, with low error rates ($10^{-6}$) and high cycling capabilities $>10^{12}$. With these perspectives, SOT-MRAM applications are primarily oriented towards replacing high-performance and high-density SRAM families. They are expected to cover registers to the L1-L3 level in central processing units and graphical processing units - while keeping in mind that commercial deployment of SOT-MRAM will rely on meeting requirements that are specific to a given technology.

Therefore, it is essential to consider the overall performances at the single-cell level and at the array size level, which includes the access lines resistance and capacitances parasitic of the control and sensing peripheries. Because the latter is extremely dependent on CMOS node and on end-user targets, assessing the potential of SOT-MRAM with proper benchmarking against SRAM requires compact device models built from large-statistics experimental data and system-design power-performance-area-cost analysis. The main difficulty of this assessment is that today's SOT-MRAM is still in the early development phase; the outcomes are extremely dependent on SOT cell performances assumptions. In addition, the work carried out on SOT-RAM design is still very limited compared to STT-MRAM. The outcomes are extremely dependent on SOT cell performances assumptions. The future of SOT-MRAM manufacturability covers etch challenges, stack performances for write, read, and retention optimization, and bit/array-cell design for density and system level. In the following, we provide a first-level analysis and project the potential of SOT-MRAM based on known silicon data and assumed material optimization.

*Integration challenges:* SOT- and STT-MRAMs share the same technological platform, with two additional modules for SOT, such as forming and aligning the SOT channel. Typically, the main integration process steps consist of four main modules [9]: i. via patterning to contact SOT track followed by SOT-MTJ stack deposition, ii. MTJ pillar definition using advanced lithography, e.g. 193 nm deep immersion, followed by ion beam etching and encapsulation, iii. SOT track definition module followed by ion beam etching, iv. a Cu dual damascene process to interconnect top and bottom contacts for routing and testing. Each of these modules is typically followed by an oxide refill and chemical-mechanical planarization to flatten the surface at nm level, as well as open masks to align the different steps. Therefore, SOT can benefit from the established process learned from the development in foundries of STT-MRAM that are affecting both the device yield and device performances. They include high quality and homogeneous stack growth, morphology and etch impact on magnetic properties and TMR, and diffusion barriers to reach back-end-of-line thermal budget (400 °C). Despite this advantage, SOT-MRAM is facing process challenges related to its three-terminal geometry. In STT, the free layer generally sits on top of the MTJ stack (bottom pinned stacks), and an overreach in oxide allows for good MgO sidewall cleaning limiting risks of defective (shorted) devices. In the case of SOT, the free layer is at the bottom (top pinned stack), and the etching of the MTJ has to be precise enough to stop on the thin SOT channel (typically < 5 nm) without degrading its conductivity nor causing vertical shorts due to metal re-deposition on tunnel barrier sidewalls [9], [491]. Meanwhile, scaling the density (typically targeting pitch < 100 nm, MTJ ≈ 30 nm) tends to enhance the risk of metal re-deposition due to shadowing effects, calling for specific SOT-MTJ etch module development. Regarding the SOT track etch module, the major challenge is the alignment precision (ideally an auto-alignment) and the width minimization with respect to MTJ diameter in order to maximize the write efficiency and minimize the total charge current.

*Performance challenges:* At the beginning of the paper, we have provided some targets for future SOT-MRAM in Fig. 1. Here, we address the challenges in realizing these targets. The read and write latencies (including delay time in access lines) should be ≈ 0.5 ns to 4 ns for the typical target applications (high performance vs. high density). While the writing speed is largely matched, the latency is affected by density and routing parasitic effects. It will require low writing currents (< 100 µA), and notably low access lines resistance, which makes it challenging to achieve the lowest technology nodes [10]. The reading latency is directly linked to the minimum current detectable by the periphery sense amplifiers, on/off ratio (TMR target > 150 %), resistance-area (RA ≈ 4 $\Omega\mu m^2$) product, and cell design (capacitance of the read bit line).

In addition, any technology suitable for embedded memory applications must be compatible with an advanced back-end line processing of CMOS chips, which typically requires deposition of the low-k dielectric at 400 °C for times varying from 30 min to hours. Hence, reaching these performances will demand strong stack engineering development efforts to reach a 400 °C thermal budget while maintaining sufficient thermal stability of the free layers (from months to years) and reference layer (years) with new SOT materials (optimized SOT efficiency and conductivity). The TMR is already reaching its targets [11], [408], and should reach the typical target of 250 % by improved MTJ tunnel barrier processing methods. Such progress would minimize read error rates and periphery area.

More critically, implementing commercial SOT-MRAMs requires the reduction of the critical switching current while preserving the device functionality and speed. The most straightforward approach is downscaling. A realistic cell dimension for a sub 28 nm node would consist of an MTJ with a diameter of 30 nm to 35 nm placed on top of a nearly-equal wide SOT-line (32 nm to 40 nm). State-of-the-art perpendicular magnetic anisotropy SOT-MTJ would require a critical current of ≈ 250 µA at a nanosecond time scale with $\theta_{SOT} \approx 0.4$ [319], [324], which is still insufficient. Based on simplified critical switching current assumptions [28] combined with experimental data, increasing $\theta_{SOT} > 0.8$ is key to achieve sub 100 µA currents (Fig. 21b), calling for the introduction of new materials, as discussed in section III (Materials for SOTs). However, while any $\theta_{SOT}$ gain will be beneficial, milestones of 1.4 seem already reasonable, noting that further increase will have lower impact because of the $1/\theta_{SOT}$ scaling of critical current.

Care is required when discussing device characteristics before drawing conclusions at the circuit and especially at the application level. Researchers have explored different SOT materials for maximizing array-level read/write performance by considering the parasitic resistances and capacitances of the access lines [492]. This array-level work highlights the importance of improving $\theta_{SOT}$ at the device level. At the system level, having a cache with a larger capacity (thanks to the better density of SOT-MRAM) in a hybrid SOT-MRAM/SRAM can achieve better performance than SRAM alone [310]. Future device-technology and device-algorithm co-optimizations require further exploration.

*Design and system challenges:* In order to minimize footprint and be competitive with other technologies in terms of areal density, a 3-terminal MTJ cell should match the power performances of a single CMOS transistor in terms of voltage and current delivery. Typically, a transistor in sub 28 nm nodes can deliver from 0.7 V to 1.1 V, which is already largely achievable [11], [324], [408], and currents on the order of 150 µA. Preliminary cell size analysis shows that standard SOT-MRAM designs are moderately competitive (Fig. 21a) because of the overly large number of control terminals (5 terminals) that are dominating the cell area. However, it is possible to reduce the number of terminals in a SOT bit cell by sharing some of them through smart designs or combined physics (e.g., SOT + STT write), leading to a cell that can be 40 % denser than SRAM cells [10] (SOT-HD1). Yoda et al. [493] proposed to pattern multi-pillars on the same tracks to extend the density even further, with a bit selectivity operated by voltage-controlled magnetic anisotropy, bringing density close to that of STT (SOT-HD2), but impacting read speeds due to a larger resistance-area product. Interestingly, a mix of these designs could be combined on the same die, such as implementing high density and high-performance computing based on only SOT (and STT) MRAM.

Based on such assumptions, system-level evaluation at 5 nm node [10] shows that read/write in SOT-MRAM becomes more energy-efficient as compared to HP-SRAM at 0.4 MB (max L1 capacity). It crosses HD-SRAM at 2 MB/8 MB for read/write, due to the exponential increase of SRAM standby power with increasing capacity (Fig. 21c). In line with some foundries stopping the scaling race, working at higher technology node (typically 22 nm) would be more beneficial, as SOT-MTJ dimension could remain similar, but routing parasitics would need to be consequently reduced. One can expect that with recent large scale processing developed by companies and R&D institutes [9], [11], [408], including co-integration with CMOS, more demonstration on the potential and performances of SOT for replacing SRAM will be shown in the near future.

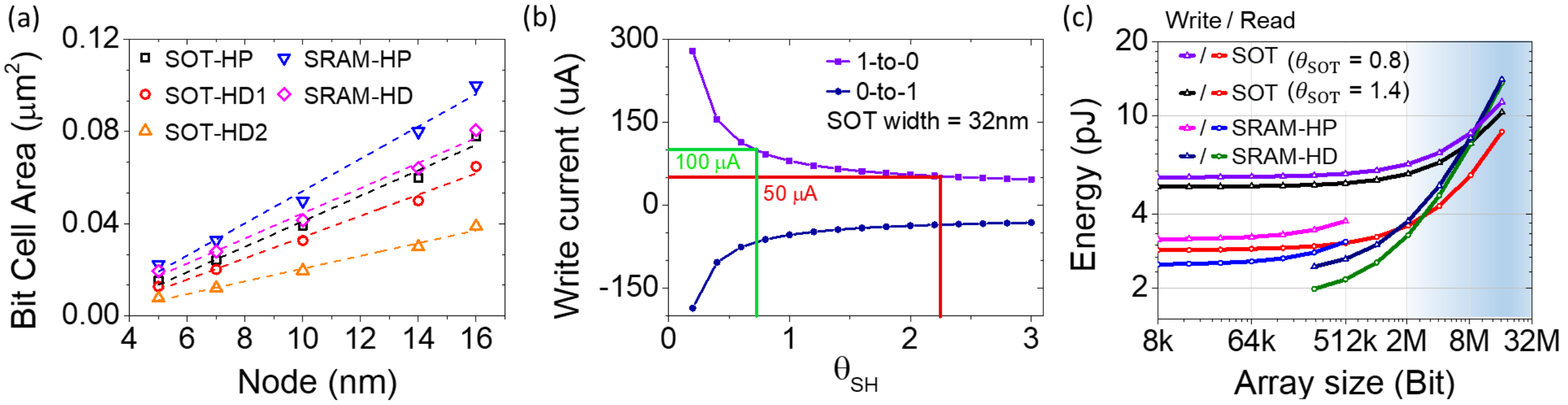

**Figure 21**. a) Bit cell area projection based on averaged foundries specs (SRAM), for three class of SOT-MTJ design targeting high density (HD) and high performance (HP) replacement of SRAM, b) critical switching current scaling as a function of spin all angle for a 32 nm width SOT track, 4 nm thick and MTJ with $B_K$ = 0.3 T, $B_X$ = 30 mT, c) Energy consumption including periphery vs. array size at 5 nm node benchmarked against SRAM.

## 5. Conclusion

Overall, spin-orbit torques have proven to be an exciting new opportunity for efficient electrical control of magnetization states. We list a few points of historical and potential developments in Fig. 22. They present a surprisingly rich array of fundamental physics that encompasses bulk, interface, intrinsic and extrinsic phenomena. At the same time, their basic properties are deeply rooted in the symmetry properties of the materials and systems. The materials palette for generating spin-orbit torques is rich and varied. It includes elemental metals and simple alloys, but it also builds on a wide variety of quantum materials, including topological insulators, Weyl semimetals, and two-dimensional van der Waals materials. At the same time, even magnetically ordered materials provide new perspectives for spin current generation. Just as varied as the available materials is the range of potential applications. The most immediate impact will most likely be for solid-state memory devices for conventional architectures and novel computational paradigms, which value a colocation of logic and data storage functionalities. But even more, unusual applications may include the generation and detection of electromagnetic radiation ranging from microwaves up to the THz spectrum. If the past decade of explosive research investigations are an indication of future scientific and engineering developments, then we can expect that spin-orbit coupling phenomena will continue to transform spintronic applications for sustainable new technologies.

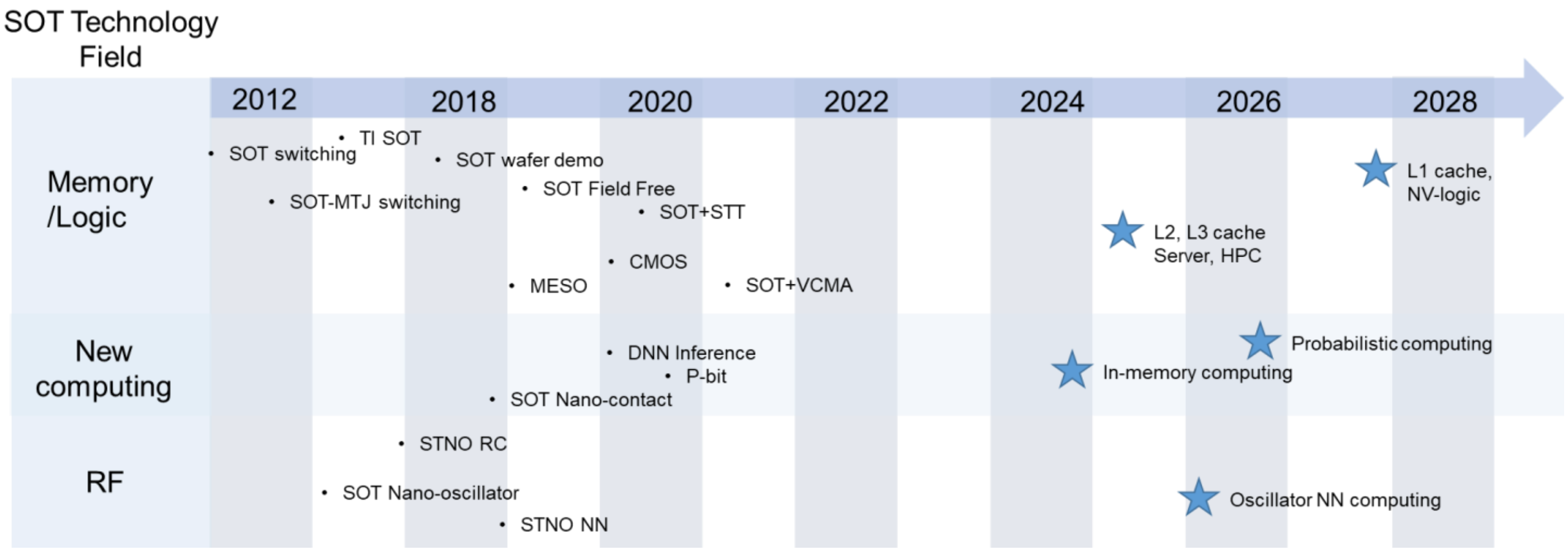

**Figure 22**. Historical and potential developments in three SOT technology fields: memory/logic, new computing, and RF (radio frequency). MESO: magneto-electric spin-orbit (logic) [494]; VCMA: voltage-controlled magnetic anisotropy; HPC: high-performance computing; NV-logic: nonvolatile-logic; P-bit: probabilistic bit; RC: reservoir computing.

## Acknowledgment

Q.S., P. L., and W.Z. coordinated this roadmap. Work on the manuscript preparation by W.Z. was supported by the National Science Foundation under award ECCS-1941426. Work on the manuscript preparation by A.H. was supported as part of Quantum Materials for Energy Efficient Neuromorphic Computing, an Energy Frontier Research Center funded by the U.S. DOE, Office of Science, under Award #DE-SC0019273. H.Y. is supported by AME-IRG through RIE2020 funds under Grant A1983c0037 and NUS Hybrid-Integrated Flexible Electronic Systems Program. Q. S. is supported by Hong Kong Research Grants Council-Early Career Scheme (Grant No. 26200520). L.L. is supported by National Science Foundation under award ECCS-1808826. Y. M. and F. F. acknowledge funding from Deutsche Forschungsgemeinschaft (DFG, German Research Foundation) - TRR 173 - 268565370 (project A11). S.F. is supported by JSPS Kakenhi 19H05622 and JST-CREST JPMJCR19K3. K.G. was supported by IMEC's Industrialization Affiliation Program on MRAM devices. We thank Laith Alahmed for his careful reading of the manuscript.